\documentclass[prd,aps,twocolumn,a4paper,showkeys,nofootinbib,10pt]{revtex4-1}

\usepackage{amsmath}
\usepackage{amsfonts}
\usepackage{amssymb}	
\usepackage{graphicx}
\usepackage{bm}
\usepackage{hyperref}
\usepackage[T1]{fontenc}
\usepackage{xcolor}
\allowdisplaybreaks

\newcommand{\diff}{\ensuremath{\mathrm{d}}}

\newcommand{\orcid}[1]{\href{https://orcid.org/#1}{
	\includegraphics[width=10pt]{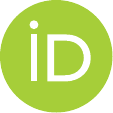}
}}

\begin{document}
\title{One-cycle radiation-reaction Magnusian for eccentric binaries through relative 1PN order}

\author{Andrea Placidi\orcid{0000-0001-8032-4416}}

\affiliation{Dipartimento di Fisica e Geologia, Universit\`a di Perugia,
	I.N.F.N. Sezione di Perugia, \\ Via Pascoli, I-06123 Perugia, Italy}

\email{andrea.placidi@unipg.it}

\begin{abstract}
	We derive the one-cycle first Magnusian for nonspinning compact
	binaries on planar, bound eccentric orbits through relative first
	post-Newtonian (1PN) order and linear order in radiation reaction, and
	construct the associated first-order periastron return map. Combining the conservative 1PN dynamics and its Jacobi transport with the
	2.5PN and 3.5PN radiation-reaction forces, we obtain all four components in
	closed form for \(0<e<1\), without expanding in eccentricity. The dependence on all eight Iyer--Will radiation-reaction gauge parameters reduces to periodic coboundaries and
	therefore cancels over a radial cycle. The result reproduces the averaged
	energy and angular-momentum losses, their circular limit, and the secular
	evolution of the 1PN radial orbital elements. The two remaining components encode the dissipative angle-sector information.
	Projecting the resulting phase-space update onto the perturbed periastron
	return section yields the corrections to the return time and apsidal phase. We then compare the strict first-order map with a
	partially exponentiated map built from the same Magnusian. The comparison is
	observable dependent: at fixed periastron index, the strict map gives smaller
	residuals for a representative evolution over 40 radial cycles, whereas
	partial exponentiation reduces the accumulated periastron-timing residual by
	about an order of magnitude. Across a representative one-cycle grid at equal masses, partial exponentiation
	improves the radial-action and energy updates at every sampled point.
\end{abstract}

\maketitle

\section{Introduction}
\label{sec:intro}

The post-Newtonian (PN) description of compact-binary dynamics separates the
conservative orbital motion from the dissipative effects produced by the
emission of gravitational radiation. For nonspinning binaries, the leading
radiation-reaction force enters at 2.5PN order and its first relative-PN
correction at 3.5PN order
\cite{Iyer:1993xi,Iyer:1995rn,Blanchet:1996vx,Pati:2002ux,
Konigsdorffer:2003ue,Nissanke:2004er,Blanchet:2013haa}. The
radiative fluxes measured at infinity are gauge invariant, whereas the local
radiation-reaction force, the mechanical losses, and the associated Schott
terms depend on the phase-space coordinates and on the radiation-reaction
gauge \cite{Iyer:1993xi,Iyer:1995rn,Blanchet:1996vx,Bini:2012ji}. The balance construction of Iyer and Will makes this freedom explicit,
and its extension through 3.5PN order involves a larger family of gauge
parameters \cite{Iyer:1993xi,Iyer:1995rn,Bini:2026suo}. A change of
radiation-reaction gauge can redistribute contributions between the local
mechanical losses and total time derivatives of Schott terms. The
instantaneous force alone therefore does not provide a gauge-independent
characterization of the dissipative evolution. It is instead natural to
consider the net evolution between equivalent points of a bound orbit, such
as two successive periastron passages, for which periodic endpoint
contributions can cancel.

For eccentric binaries, the standard orbit-averaged balance equations provide
the secular evolution of the orbital energy, angular momentum, and radial
elements \cite{Peters:1963ux,Peters:1964zz,Wagoner:1976am,
Blanchet:1989cu,Junker:1992kle,Gopakumar:1997bs,Damour:2004bz,
Konigsdorffer:2006zt,Arun:2007rg,Arun:2007sg,Arun:2009mc}. They do not, however, directly organize the complete phase-space evolution over one radial cycle, including the shifts of the variables conjugate to the orbital actions. A finite-time generator provides a natural way of encoding this additional information \cite{Kim:2025gis}. In the conservative case, finite-time evolution can be organized with the Magnus expansion \cite{Magnus:1954zz,Blanes:2008xlr}. For a dissipative system, an ordinary Hamiltonian generator on the physical phase space is unavailable, so a different canonical representation is required.

Blanco recently extended the Magnusian construction to dissipative and
nonlocal-in-time dynamics \cite{Blanco:2026prj,Blanco:2024fte} by embedding the physical
evolution in the doubled phase space of the classical in-in formalism
\cite{Galley:2012hx,Galley:2009px,Galley:2014wla}. At first order in the
perturbing force, the resulting Magnusian is obtained by evaluating the force
along the unperturbed orbit and transporting it back to the initial
phase-space point with the Jacobi propagator. Blanco then applied this formula
to Newtonian bound motion with the leading 2.5PN radiation-reaction force and
derived the corresponding one-cycle map without expanding in eccentricity. We take
that general construction as our starting point. In particular, we do not
repeat the doubled-phase-space derivation, for which we refer to
Ref.~\cite{Blanco:2026prj}, and concentrate instead on extending its
compact-binary application.

The extension through relative 1PN order is not obtained by simply evaluating
the 3.5PN force on a Newtonian ellipse. At the same accuracy one must include
the 1PN conservative Hamiltonian, the resulting corrections to the
unperturbed flow and to its Jacobi propagator, the transformation between the
coordinates in which the radiation-reaction force is given and a common
canonical phase space, and the direct 3.5PN contribution to the force. These
ingredients must then be combined consistently over one radial cycle. We
perform this calculation in normal-form Delaunay variables, construct their
explicit canonical embedding into Arnowitt--Deser--Misner (ADM) phase space,
and do not expand in eccentricity.

Our main result is the complete four-component first Magnusian through
relative 1PN order for planar, eccentric, nonspinning binaries. Starting from
the full Iyer--Will gauge family, we show that its parameter dependence is a
periodic coboundary whose endpoint contribution vanishes over a radial cycle.
The final one-cycle generator is therefore independent of all eight
Iyer--Will gauge parameters once the phase-space variables are transformed
consistently. At leading order we recover the result of
Ref.~\cite{Blanco:2026prj}.
We further distinguish the one-cycle Magnusian, constructed over the fixed
conservative radial interval, from the physical periastron return map.
Projecting the fixed-time endpoint onto the perturbed return section yields
the first-order correction to the periastron arrival time and, at relative
1PN order, an additional dissipative correction to the apsidal phase.

To our knowledge, the relative-1PN contribution, the treatment of the full
eight-parameter Iyer--Will family, and the associated physical checks have not
appeared previously in the literature.

We verify the result analytically through energy and angular-momentum balance,
the circular limit, the secular evolution of the radial orbital elements, and
the shear structure of the conservative normal-form flow. We then use the
same shear structure to construct the projection onto the periastron return
section and validate the resulting timing and apsidal corrections against the
direct dynamics.

We also compare the strict first-order map with the partially exponentiated
prescription of Ref.~\cite{Blanco:2026prj}. Their relative performance is
observable dependent: for the representative long evolution, the strict map
better tracks the state at fixed periastron index, whereas partial
exponentiation substantially reduces the accumulated timing error. A
complementary one-cycle grid shows that partial exponentiation improves the
radial-action and energy updates, the strict map improves the angular-action
update, and the eccentricity comparison is mixed.

The paper is organized as follows. Sec.~\ref{sec:construction} summarizes
the doubled-phase-space construction and specializes the first Magnusian to a
radial cycle. Sec.~\ref{sec:conservative-normal-form} builds the conservative
1PN normal form, ADM embedding, and Jacobi transport. The closed-form result is
derived in Sec.~\ref{sec:one-cycle-result}. Its balance and structural checks,
together with the projection onto the periastron return section, are discussed
in Sec.~\ref{sec:physical-checks}.
Sec.~\ref{sec:gauge} establishes independence from all eight Iyer--Will gauge
parameters. Sec.~\ref{sec:blanco-reduction} compares the leading-order result
with Ref.~\cite{Blanco:2026prj} and relates the two map prescriptions.
Sec.~\ref{sec:numerical-comparison} then presents their numerical comparison.
Sec.~\ref{sec:conclusions} summarizes our conclusions and outlook. The
periodic-coboundary proof and the formal radiation-reaction-strength
diagnostic are given in Appendices~\ref{app:periodic-coboundary} and
\ref{app:formal-lambda-diagnostic}, respectively.

\subsection{Conventions and perturbative counting}
\label{subsec:notation}

We consider two nonspinning bodies of masses $m_1$ and $m_2$, and define
\begin{equation}
	M\equiv m_1+m_2,\qquad
	\mu\equiv\frac{m_1m_2}{M},\qquad
	\nu\equiv\frac{\mu}{M}.
\end{equation}
We use the reduced Hamiltonian $H\equiv H_{\rm physical}/\mu$ and reduced canonical momenta, and
measure lengths and times in units of the total mass. Unless explicitly
restored, $G_{\rm N}=c=M=1$, so that the only remaining mass parameter is the
symmetric mass ratio $0<\nu\leq 1/4$.

The planar dynamics is described initially by ADM polar canonical variables
\begin{equation} \label{eq:adm_can_var}
	\bm z\equiv(r,\phi,p_r,p_\phi),
	\qquad
	\{r,p_r\}=\{\phi,p_\phi\}=1.
\end{equation}
For the conservative normal form we use Delaunay variables
\begin{equation}
	\bm Y
	\equiv
	(\ell,g,\mathcal L,\mathcal G),
	\qquad
	\{\ell,\mathcal L\}
	=
	\{g,\mathcal G\}
	=
	1.
\end{equation}
Here \(\ell\) is the mean anomaly and \(g\) is the argument of periastron.
Their conjugate variables \(\mathcal L\) and \(\mathcal G\) are,
respectively, the principal Delaunay action and the angular-momentum action.
At Newtonian order, \(\mathcal L\) determines the semimajor axis and the
binding energy, whereas \(\mathcal G\) equals the reduced orbital angular
momentum \(p_\phi\). Explicitly,
\begin{align}
	a
	&=
	\mathcal L^2,
	&
	\mathcal G
	&=
	p_\phi
	=
	\mathcal L\sqrt{1-e^2},
	\nonumber\\
	H_{\rm N}
	&=
	-\frac{1}{2\mathcal L^2},
	&
	n
	&=
	\mathcal L^{-3},
\end{align}
where \(a\) is the semimajor axis, \(e\) the eccentricity, and \(n\) the mean motion.
We will continue to use the quantity
\begin{equation} \label{eq:ecc_param}
	e\equiv\sqrt{1-\frac{\mathcal G^2}{\mathcal L^2}}
\end{equation}
as an action-based eccentricity parameter throughout the 1PN analysis. It should not be confused with the 1PN radial eccentricity
$e_r$, which
will be introduced when discussing the physical radial elements.

 The Delaunay variables are well defined for \(0<e<1\). At \(e=0\), the
argument of periastron \(g\) is undefined, so the angles \(\ell\) and \(g\)
no longer have a separate meaning; the regular circular phase is their sum
\(\ell+g\). Whenever needed, the circular limit is therefore taken by
continuity from \(e>0\).


We also introduce two formal bookkeeping parameters. The parameter
\(\varepsilon\) counts conservative PN orders, whereas \(\lambda\) counts
powers of the radiation-reaction interaction (with the \(k\)th Magnusian correspondingly scaling as \(\lambda^k\)):
\begin{align}
	H_{\rm cons}
	&=H_{\rm N}+\varepsilon H_{\rm 1PN}
	+\mathcal O(\varepsilon^2),\\
	\mathcal F_{\rm RR}
	&=\lambda\left(
	\mathcal F_{\rm 2.5PN}
	+\varepsilon\mathcal F_{\rm 3.5PN}
	\right)
	+\mathcal O(\lambda\varepsilon^2).
\end{align}
The parameters $\varepsilon$ and $\lambda$ are treated as formally independent in the
calculation. In the physical PN counting, however, the leading
radiation-reaction force is of 2.5PN order, so that one may assign $\lambda\sim\varepsilon^{5/2}$.
Consequently, the terms proportional to \(\lambda\) and
\(\lambda\varepsilon\) correspond respectively to the 2.5PN and 3.5PN
radiation-reaction contributions. The terms linear in the force but omitted from our
calculation begin at 4.5PN,
\(\lambda\varepsilon^2\sim\varepsilon^{9/2}\). Terms quadratic in the
radiation-reaction strength would instead begin at 5PN, since
\(\lambda^2\sim\varepsilon^5\).

The full Magnusian admits the expansion
\begin{equation}
	\chi_{\rm M}
	=
	\sum_{k\geq1}\lambda^k\chi^{(k)},
	\label{eq:magnus-expansion}
\end{equation}
where $\chi^{(k)}$ denotes the $k$th Magnus coefficient. In this work we determine $\chi^{(1)}$ through relative 1PN
order,
\begin{equation}
	\chi^{(1)}
	=
	\chi_{\rm N}^{(1)}
	+
	\varepsilon\chi_{\rm 1PN}^{(1)}
	+
	\mathcal O(\varepsilon^2).
	\label{eq:first-magnusian-pn-expansion}
\end{equation}
The leading term $\chi_{\rm N}^{(1)}$ is generated by the 2.5PN
radiation-reaction force evaluated on the Newtonian conservative dynamics,
whereas $\chi_{\rm 1PN}^{(1)}$ contains both the direct 3.5PN contribution
and the relative-1PN corrections induced by the conservative dynamics.
Higher Magnus coefficients, beginning with $\chi^{(2)}$, are beyond the
scope of the present calculation.
\section{First-order one-cycle construction}
\label{sec:construction}

\subsection{Interaction-picture Magnusian}
\label{subsec:interaction-picture}

The reduced equations of motion contain a conservative Hamiltonian flow and a
non-Hamiltonian radiation-reaction contribution. Following the classical
in-in construction of Refs.~\cite{Galley:2012hx,Galley:2009px,Galley:2012qs,
Galley:2014wla},
the latter can be embedded
in a canonical dynamical system on a doubled phase space. We shall not repeat that
construction here, but only summarize the first-order result needed below;
for its derivation and its extension to nonlocal-in-time forces, we refer to
Ref.~\cite{Blanco:2026prj}.

Let $Q_1^A$ and $Q_2^A$ denote the two auxiliary phase-space histories of the
in-in construction. We introduce the average and difference variables
\begin{equation}
	Q_+^A
	\equiv
	\frac{Q_1^A+Q_2^A}{2},
	\qquad
	Q_-^A
	\equiv
	Q_1^A-Q_2^A,
	\label{eq:plus-minus-variables}
\end{equation}
where the index $A$ runs over both coordinates and momenta. The two histories
are auxiliary: the physical dynamics is recovered by taking the physical
limit (PL)
\begin{equation}
	Q_-^A\rightarrow0,
	\qquad
	Q_+^A\rightarrow Q^A,
	\label{eq:physical-limit}
\end{equation}
after the doubled evolution has been generated.

Let the ordinary phase-space Poisson bracket be defined by
$\{Q^A,Q^B\}=\Omega^{AB}$. The doubled bracket is
\begin{equation}
	\{\cdot,\cdot\}_{\rm d}
	\equiv
	\{\cdot,\cdot\}_1-\{\cdot,\cdot\}_2.
	\label{eq:doubled-bracket-definition}
\end{equation}
In terms of the variables in Eq.~\eqref{eq:plus-minus-variables}, it satisfies
\begin{subequations}
	\begin{align}
		\{Q_+^A,Q_+^B\}_{\rm d}
		&=
		\{Q_-^A,Q_-^B\}_{\rm d}
		=0,
		\\
		\{Q_+^A,Q_-^B\}_{\rm d}
		&=
		\Omega^{AB}.
		\label{eq:doubled-poisson-brackets}
	\end{align}
\end{subequations}
Thus, the average and difference variables form cross-conjugate sectors of
the doubled phase space.

Let $X^{(0)A}_{t,t_i}(Q)$ be the conservative flow from $t_i$ to $t$, and
define its tangent, or Jacobi, propagator by
\begin{equation}
	\mathsf M^A{}_{B}(t,t_i;Q)
	\equiv
	\frac{\partial X^{(0)A}_{t,t_i}(Q)}
	{\partial Q^B}.
	\label{eq:jacobi-general}
\end{equation}
Let $R_{\rm RR}^A$ denote the dissipative vector field added to the
conservative Hamiltonian equations. The covector entering the doubled
interaction Hamiltonian is its symplectic lowering,
\begin{equation}
	\mathfrak F_A
	\equiv
	\Omega_{AB}R_{\rm RR}^B,
	\qquad
	H_I(Q_+,Q_-)
	=
	Q_-^A\mathfrak F_A(Q_+).
	\label{eq:dissipative-covector-lowering}
\end{equation}
At first order in
$\lambda$, the interaction-picture Magnusian is given by
\begin{equation}
	\lambda \, \chi^{(1)}
	=
	-Q_-^B
	\int_{t_i}^{t_f}\!\diff t\,
	\left[
	\left(X^{(0)}_{t,t_i}\right)^*\mathfrak F
	\right]_B(Q_+),
	\label{eq:first-magnusian-general}
\end{equation}
where
\begin{equation}
	\mathfrak F
	\equiv
	\mathfrak F_A(Q)\,\diff Q^A,
	\qquad
	\mathfrak F_A=\mathcal O(\lambda),
	\label{eq:force-one-form-general}
\end{equation}
is the doubled-space force covector. Its pullback along the conservative flow has
components
\begin{equation}
	\left[
	\left(X^{(0)}_{t,t_i}\right)^*\mathfrak F
	\right]_B(Q)
	=
	\mathfrak F_A
	\left[
	X^{(0)}_{t,t_i}(Q)
	\right]
	\mathsf M^A{}_{B}(t,t_i;Q).
	\label{eq:force-pullback}
\end{equation}
Eq.~\eqref{eq:first-magnusian-general} therefore evaluates the force
covector along the conservative orbit and pulls it back to the initial
phase-space point. Indeed, an initial variation $\delta Q^B$ is transported
by the conservative flow according to
$\delta X^{(0)A}(t)=\mathsf M^A{}_{B}(t,t_i;Q)\,\delta Q^B$. Consequently, the contraction of the force covector with the transported variation can be written as
\begin{widetext}
	\begin{align}
		\mathfrak F_A\!\left[X^{(0)}_{t,t_i}(Q)\right]
		\delta X^{(0)A}(t)
		=
		\mathfrak F_A\!\left[X^{(0)}_{t,t_i}(Q)\right]
		\mathsf M^A{}_{B}(t,t_i;Q)\,\delta Q^B,
	\end{align}
\end{widetext}
showing that
$\mathfrak F_A[X^{(0)}] \mathsf M^A{}_{B}$ are precisely the components of the force covector pulled back to the initial phase-space point.

Following Refs.~\cite{Magnus:1954zz,Blanes:2008xlr}, the formal finite map generated by the complete Magnus series is
\begin{equation}
	\mathcal O_f
	=
	\exp\!\left(
	\lambda\mathcal D_1+
	\lambda^2\mathcal D_2+\cdots
	\right)\mathcal O_i,
	\label{eq:formal-magnus-exponential}
\end{equation}
where 
\begin{equation}
	\mathcal D_k\mathcal O
	\equiv
	\left.
	\left\{
	\chi^{(k)},\mathcal O
	\right\}_{\rm d}
	\right|_{\rm PL}.
	\label{eq:magnus-derivation-definition}
\end{equation}
This is equivalent to the Lie-series notation used in
Ref.~\cite{Blanco:2026prj}.\footnote{Given the complete Magnusian of Eq.~\eqref{eq:magnus-expansion}, one has
	\(\{\chi_{\rm M},*\}_{\rm d}|_{\rm PL}
	=\sum_{k\geq1}\lambda^k\mathcal D_k\). Therefore,
	Eq.~\eqref{eq:formal-magnus-exponential} is equivalently written as
	\(\mathcal O_f
	=\exp\!\bigl(\{\chi_{\rm M},*\}_{\rm d}|_{\rm PL}\bigr)\mathcal O_i\),
	which is the form displayed in Ref.~\cite{Blanco:2026prj}.}
Expanding Eq.~\eqref{eq:formal-magnus-exponential} consistently through first
order gives
\begin{align}
	\mathcal O_f
	&=
	\mathcal O_i
	+
	\lambda\mathcal D_1\mathcal O_i
	+
	\mathcal O(\lambda^2)
	\nonumber\\
	&=
	\mathcal O_i
	+
	\lambda\left.
	\left\{
	\chi^{(1)},\mathcal O_i
	\right\}_{\rm d}
	\right|_{\rm PL}
	+
	\mathcal O(\lambda^2).
	\label{eq:first-order-observable-map}
\end{align}
We refer to Eq.~\eqref{eq:first-order-observable-map} as the strict
first-order map. It is the perturbatively consistent truncation of the same
exponential used in Ref.~\cite{Blanco:2026prj}.
In that prescription, partial exponentiation retains selected nested actions
generated by $\mathcal D_1$ beyond linear order. We compare the two
truncations in Secs.~\ref{sec:blanco-reduction} and
\ref{sec:numerical-comparison}.

It is convenient to define the covector associated with the complete first-order term $\lambda\chi^{(1)}$ by~\footnote{With this definition, the covector $\chi_B^{(1)}$ carries an explicit factor of $\lambda$, whereas the scalar Magnus coefficient $\chi^{(1)}$ itself remains independent of $\lambda$.}

\begin{equation} \label{eq:covector_def}
\lambda\chi^{(1)}
	\equiv
	Q_-^B\chi_B^{(1)},
\end{equation}
according to which
\begin{align}
	\chi_B^{(1)}
	&=
	-\int_{t_i}^{t_f}\!\diff t\,
	\left[
	\left(X^{(0)}_{t,t_i}\right)^*\mathfrak F
	\right]_B(Q_+)
	\nonumber\\
	&=
	-\int_{t_i}^{t_f}\!\diff t\,
	\mathfrak F_A
	\left[
	X^{(0)}_{t,t_i}(Q_+)
	\right]
	\mathsf M^A{}_{B}(t,t_i;Q_+).
	\label{eq:magnusian-coefficient-covector}
\end{align}
Applying this relation to the phase-space coordinates gives
\begin{equation}
	\Delta Q^A
	=
	-\Omega^{AB}\chi^{(1)}_B
	+
	\mathcal O(\lambda^2).
	\label{eq:physical-kick-general}
\end{equation}

Specializing Eq.~\eqref{eq:physical-kick-general} to the Delaunay ordering
\begin{equation}
	Q^A=(\ell,g,\mathcal L,\mathcal G),
	\quad
	\chi_A^{(1)}
	\equiv
	\bigl(
	\chi_\ell^{(1)},\chi_g^{(1)},
	\chi_{\mathcal L}^{(1)},\chi_{\mathcal G}^{(1)}
	\bigr),
\end{equation}
the strict first-order interaction-picture kick associated with one radial
cycle is
\begin{align}
	\bigl(\Delta\ell,\Delta g\bigr)_{\rm RR}
	&=
	\bigl(-\chi_{\mathcal L}^{(1)},-\chi_{\mathcal G}^{(1)}\bigr)
	+
	\mathcal O(\lambda^2),
	\\
	\bigl(\Delta\mathcal L,\Delta\mathcal G\bigr)_{\rm RR}
	&=
	\bigl(\chi_\ell^{(1)},\chi_g^{(1)}\bigr)
	+
	\mathcal O(\lambda^2).
	\label{eq:delaunay-kick-from-chi}
\end{align}
Thus, $\chi_\ell^{(1)}$ and $\chi_g^{(1)}$ give the one-cycle
changes of the two actions, whereas $-\chi_{\mathcal L}^{(1)}$ and
$-\chi_{\mathcal G}^{(1)}$ give the corresponding interaction-picture
angle kicks. The latter encode radial-phase and apsidal information that is
not determined by the usual balance laws. They should not, however, be
identified directly with the physical angle shifts on the next periastron
section. The relation between the interaction-picture kick and the physical
return-section quantities is derived in
Sec.~\ref{subsec:return-section}.

Returning to the main goal of our calculation, if we expand the coefficient
covector of the first Magnusian as
\begin{equation} \label{eq:chi1_exp_eps}
	\chi_A^{(1)}
	=
	\chi_{{\rm N},A}^{(1)}
	+
	\varepsilon\chi_{{\rm 1PN},A}^{(1)}
	+
	\mathcal O( \varepsilon^2),
\end{equation}
the main task of the present work is to determine the
relative-1PN coefficient $\chi_{{\rm 1PN},A}^{(1)}$.

\subsection{Post-Newtonian dynamical input}
\label{subsec:pn-input}

We work in canonical ADM polar variables. Through 1PN order, the reduced
conservative Hamiltonian is
\begin{equation}
	H_{\rm cons}
	=
	H_{\rm N}
	+
	\varepsilon H_{\rm 1PN}^{\rm ADM}
	+
	\mathcal O(\varepsilon^2),
	\label{eq:adm-hamiltonian-expansion}
\end{equation}
where
\begin{subequations}
	\label{eq:H_PN_coeffs}
	\begin{align}
		&H_{\rm N}
		=
		\frac{p^2}{2}-\frac{1}{r},
		\label{eq:newtonian-hamiltonian}\\
		&H_{\rm 1PN}^{\rm ADM}
		=
		\frac{3\nu-1}{8}\left(p^2\right)^2
		-\frac{1}{2r}
		\left[
		(3+\nu)p^2+\nu p_r^2
		\right]
		\nonumber\\
		&\quad
		+\frac{1}{2r^2},
		\label{eq:1pn-adm-hamiltonian}
	\end{align}
\end{subequations}
and $p^2=p_r^2+p_\phi^2/r^2$.
These conventions agree with the standard reduced ADM dynamics reviewed, for
example, in Refs.~\cite{Blanchet:2013haa,Schafer:2018jfw}.

The dissipative equations of motion are written as
\begin{align}
	\dot r
	&=
	\frac{\partial H_{\rm cons}}{\partial p_r},
	&
	\dot\phi
	&=
	\frac{\partial H_{\rm cons}}{\partial p_\phi},
	\nonumber\\
	\dot p_r
	&=
	-\frac{\partial H_{\rm cons}}{\partial r}
	+\mathcal F_r,
	&
	\dot p_\phi
	&=
	\mathcal F_\phi.
	\label{eq:adm-dissipative-eom}
\end{align}
In the ADM ordering \(\bm z=(r,\phi,p_r,p_\phi)\), the dissipative vector
field is \(R_{\rm RR}^A=(0,0,\mathcal F_r,\mathcal F_\phi)\). Equation
\eqref{eq:dissipative-covector-lowering} therefore gives the doubled-space
force covector
\begin{equation}
	\mathfrak F_{\rm RR}
	=
	-\mathcal F_r\,\diff r-\mathcal F_\phi\,\diff\phi.
	\label{eq:polar-force-one-form}
\end{equation}
The minus signs are a consequence of symplectic lowering; the physical
generalized forces \(\mathcal F_r\) and \(\mathcal F_\phi\) themselves enter
the momentum equations with the signs displayed in
Eq.~\eqref{eq:adm-dissipative-eom}.

A convenient parametrization of the reduced radiation-reaction force is
\begin{align}
	&\bm F_{\rm RR}
	=
	\lambda\frac{8\nu}{5r^3}
	\Big\{
	\left[
	A_{\rm 2.5PN}
	+\varepsilon A_{\rm 3.5PN}
	\right]p_r\bm n
	\\\quad&-
	\left[
	B_{\rm 2.5PN}
	+\varepsilon B_{\rm 3.5PN}
	\right]\bm p
	\Big\}
	+
	\mathcal O(\lambda\varepsilon^2),
	\label{eq:rr-force-decomposition}
\end{align}
where $\bm n$ is the radial unit vector and
\begin{equation}
	\bm p
	=
	p_r\bm n+\frac{p_\phi}{r}\bm e_\phi.
\end{equation}
The corresponding generalized polar forces are
\begin{equation}
	\mathcal F_r
	=
	\bm n\cdot\bm F_{\rm RR},
	\qquad
	\mathcal F_\phi
	=
	r\,\bm e_\phi\cdot\bm F_{\rm RR}.
	\label{eq:generalized-polar-force}
\end{equation}

At leading order, the two-parameter Iyer--Will family is described by
Refs.~\cite{Iyer:1993xi,Iyer:1995rn}:
\begin{align}
	A_{\rm 2.5PN}
	&=
	3(1+\beta)p^2
	+\left(
	\frac{23}{3}+2\alpha-3\beta
	\right)\frac{1}{r}
	-5\beta p_r^2,
	\label{eq:A25}\\
	B_{\rm 2.5PN}
	&=
	(2+\alpha)p^2
	+(2-\alpha)\frac{1}{r}
	-3(1+\alpha)p_r^2.
	\label{eq:B25}
\end{align}
The harmonic-gauge choice $(\alpha,\beta)=(-1,0)$ reproduces the force used
in the leading calculation of Ref.~\cite{Blanco:2026prj}.

At 3.5PN order, $A_{\rm 3.5PN}$ and $B_{\rm 3.5PN}$ are polynomials in
$p^2$, $p_r^2$, and $1/r$. Their coefficients depend on $\nu$ and on the
six additional Iyer--Will parameters
$\delta_1,\ldots,\delta_5,\epsilon_5$
\cite{Iyer:1993xi,Iyer:1995rn,Blanchet:1996vx,Pati:2002ux,
Konigsdorffer:2003ue,Nissanke:2004er,Bini:2026suo}. For the
closed-form calculation we first use the ADM representative of this family, so
that the conservative Hamiltonian, the phase-space variables, and the
radiation-reaction force all refer to the same canonical coordinates. The
calculation is then repeated for the complete eight-parameter family in
Sec.~\ref{sec:gauge}, where its gauge response is evaluated explicitly.

\subsection{Specialization to one radial cycle}
\label{subsec:one-cycle-specialization}

We group the normal-form Delaunay variables into the angle and action vectors
\begin{equation}
	\bm\theta=(\ell,g),
	\qquad
	\bm J=(\mathcal L,\mathcal G),
	\label{eq:delaunay-vector-ordering}
\end{equation}
so that $\bm Y=(\bm\theta,\bm J)$.
Let $\overline H(\bm J)$ denote the conservative normal-form Hamiltonian,
whose explicit expression through 1PN order will be derived in the following
section. Its frequency vector is
\begin{equation}
	\bm\omega(\bm J)
	\equiv
	\frac{\partial\overline H}{\partial\bm J}
	=
	\bigl(
	\omega_\ell(\bm J),
	\omega_g(\bm J)
	\bigr),
	\label{eq:normal-form-frequencies}
\end{equation}
where $\omega_\ell$ is the radial frequency and $\omega_g$ is the
periastron-precession frequency. The normal-form flow is therefore
\begin{equation}
	\bm\theta(t)
	=
	\bm\theta_0+t\,\bm\omega(\bm J_0),
	\qquad
	\bm J(t)=\bm J_0,
	\label{eq:delaunay-normal-flow}
\end{equation}
and the radial period is
\begin{equation}
	T_r(\bm J_0)
	=
	\frac{2\pi}{\omega_\ell(\bm J_0)}.
	\label{eq:radial-period}
\end{equation}

In the ordering $(\bm\theta,\bm J)$, the Jacobi matrix of the normal-form
flow is
\begin{equation}
	\mathsf M_{\bm Y}(t)
	\equiv
	\frac{\partial\bm Y(t;\bm Y_0)}
	{\partial\bm Y_0}
	=
	\begin{pmatrix}
		\mathsf I_{2\times2} &
		t\,D_{\bm J}\bm\omega\\[1ex]
		\mathsf 0_{2\times2} &
		\mathsf I_{2\times2}
	\end{pmatrix},
	\label{eq:delaunay-jacobi}
\end{equation}
where
\begin{equation}
	\left[
	D_{\bm J}\bm\omega
	\right]_{ij}
	\equiv
	\frac{\partial\omega_i}{\partial J_j}
	=
	\frac{\partial^2\overline H}
	{\partial J_i\,\partial J_j},
	\qquad
	(J_1,J_2)=(\mathcal L,\mathcal G),
	\label{eq:frequency-jacobian}
\end{equation}
is the Hessian of the normal-form Hamiltonian in the action sector.

The normal-form Delaunay variables are convenient for describing the
conservative evolution and its Jacobi propagator, whereas the explicit
radiation-reaction force entering the Magnusian is expressed in ADM polar
variables. To evaluate this force along the conservative orbit and pull it
back to the initial Delaunay data, we therefore need the canonical map between
the two phase-space descriptions. We denote it by $\bm z
	=
	\bm Z(\bm Y)$,  where $\bm z$ is the vector of ADM canonical phase-space variables defined in Eq.~\eqref{eq:adm_can_var}.
Its explicit 1PN expression will also be constructed in the following
section.
The conservative ADM orbit generated from the initial Delaunay data
$\bm Y_0=(\ell_0,g_0,\mathcal L,\mathcal G)$ is
$\bm z^{(0)}(t;\bm Y_0)=\bm Z\bigl(\bm Y(t;\bm Y_0)\bigr)$.

The corresponding tangent map from the initial Delaunay variables to
the ADM orbit is therefore
\begin{align}
	\mathsf K^a{}_{A}(t;\bm Y_0)
	&\equiv
	\frac{\partial z^{(0)a}(t;\bm Y_0)}
	{\partial Y_0^A}
	\nonumber\\
	&=
	\left.
	\frac{\partial Z^a}{\partial Y^B}
	\right|_{\bm Y(t;\bm Y_0)}
	\left[
	\mathsf M_{\bm Y}(t)
	\right]^B{}_{A}.
	\label{eq:composed-adm-tangent-map}
\end{align}
Here and below, lower-case Latin indices $a,b$ label the ADM
variables $(r,\phi,p_r,p_\phi)$, while upper-case Latin indices $A,B$
label the Delaunay variables $(\ell,g,\mathcal L,\mathcal G)$. Indices $i,j$
label two-dimensional configuration-space components.


Using the tangent map in Eq.~\eqref{eq:composed-adm-tangent-map},
the general pullback formula \eqref{eq:magnusian-coefficient-covector}
specializes to
\begin{align}
	\chi_A^{(1)}(\bm Y_0)
	&=
	\int_0^{T_r}\!\diff t\,
	\mathcal F_r[\bm z^{(0)}(t)]
	\,\mathsf K^r{}_{A}(t)
	\nonumber\\
	&\quad
	+
	\int_0^{T_r}\!\diff t\,
	\mathcal F_\phi[\bm z^{(0)}(t)]
	\,\mathsf K^\phi{}_{A}(t)
	\nonumber\\
	&=
	\int_0^{T_r}\!\diff t\,
	\mathcal F_r[\bm z^{(0)}(t)]
	\frac{\partial r^{(0)}(t)}{\partial Y_0^A}
	\nonumber\\
	&\quad
	+
 \int_0^{T_r}\!\diff t\,
	\mathcal F_\phi[\bm z^{(0)}(t)]
	\frac{\partial\phi^{(0)}(t)}{\partial Y_0^A},
	\label{eq:one-cycle-magnusian-integral}
\end{align}
with $A\in\{\ell,g,\mathcal L,\mathcal G\}$. Thus, the physical generalized
forces evaluated along the conservative ADM orbit are contracted with the
corresponding components of the tangent map, $\mathsf K^r{}_{A}$ and
$\mathsf K^\phi{}_{A}$. The plus signs in
Eq.~\eqref{eq:one-cycle-magnusian-integral} result from combining the minus
sign in Eq.~\eqref{eq:magnusian-coefficient-covector} with the symplectic
lowering in Eq.~\eqref{eq:polar-force-one-form}.

As an immediate sign check, at Newtonian order in the circular limit one has
\(\mathcal F_\phi=-32\lambda\nu/(5\mathcal L^7)\),
\(\mathsf K^\phi{}_{\ell}=1\), and
\(T_r=2\pi\mathcal L^3\). Equation
\eqref{eq:one-cycle-magnusian-integral} then gives
\(\chi_\ell^{(1)}=-64\pi\lambda\nu/(5\mathcal L^4)\), consistently with
the action loss obtained below.

For the one-cycle Magnusian, we choose the initial point of the conservative
reference orbit at periastron, $\ell_0=0$. Rotational invariance also allows
us to set $g_0=0$, so that
$\bm Y_0=(0,0,\mathcal L,\mathcal G)$. The integration endpoint
$t=T_r(\bm J_0)$ is the next periastron of this unperturbed normal-form
orbit. In the presence of radiation reaction, the physical return to the
periastron section occurs at a slightly shifted time; this distinction is
addressed explicitly in Sec.~\ref{subsec:return-section}.
The resulting one-cycle coefficients depend only on
$(\mathcal L,\mathcal G,\nu)$, or equivalently on
$(\mathcal L,e,\nu)$. Determining the expansion in $\varepsilon$ of
Eq.~\eqref{eq:one-cycle-magnusian-integral} is
the main goal of the calculation performed below. Its leading term reproduces the known
2.5PN construction of Ref.~\cite{Blanco:2026prj}, whereas its relative-1PN
part combines the direct 3.5PN force with the 1PN corrections to the
conservative orbit, the radial period, the canonical ADM map, and the Jacobi
propagator.


\section{Conservative 1PN normal form and ADM embedding}
\label{sec:conservative-normal-form}

We now construct the conservative normal form, ADM embedding, time
parametrization, and fixed-time tangent map required by the relative-1PN
one-cycle pullback.

\subsection{Delaunay normal form}
\label{subsec:delaunay-normal-form}

We begin with the Newtonian Kepler map. For bound prograde orbits, we use the
action-based eccentricity $e$ defined in Eq.~\eqref{eq:ecc_param} and introduce
\begin{subequations}
	\begin{align}
		s
		&\equiv
		\sqrt{1-e^2}
		=
		\frac{\mathcal G}{\mathcal L},
		\\
		\rho(u)
		&\equiv
		1-e\cos u,
		\label{eq:kepler-definitions}
	\end{align}
\end{subequations}
where \(u\) is the eccentric anomaly. The mean anomaly is related to \(u\)
by Kepler's equation,
\begin{equation}
	\ell=u-e\sin u.
	\label{eq:kepler-equation}
\end{equation}
We denote by \(v(u,e)\) the continuous true anomaly, normalized by
\(v(0,e)=0\), and defined over one radial cycle by
\begin{equation}
	\frac{\partial v}{\partial u}
	=
	\frac{s}{\rho(u)},
	\qquad
	v(2\pi,e)=2\pi.
	\label{eq:true-anomaly-definition}
\end{equation}
The Newtonian canonical map from Delaunay to ADM polar variables is then
\begin{equation}
	\bm Z_{\rm N}(\bm Y)
	=
	\left(
	\mathcal L^2\rho,
	g+v,
	\frac{e\sin u}{\mathcal L\rho},
	\mathcal G
	\right).
	\label{eq:newtonian-kepler-map}
\end{equation}

Since \(\diff\ell/\diff u=\rho\), the mean-anomaly average of a function
\(A(u)\) is
\begin{equation}
	\left\langle A\right\rangle_\ell
	\equiv
	\frac{1}{2\pi}
	\int_0^{2\pi}\!\diff u\,
	\rho(u)A(u).
	\label{eq:mean-anomaly-average}
\end{equation}
Evaluating the 1PN ADM Hamiltonian in Eq.~\eqref{eq:1pn-adm-hamiltonian} on
Eq.~\eqref{eq:newtonian-kepler-map} and taking this average gives
\begin{equation}
	\overline H_{\rm 1PN}(\mathcal L,\mathcal G)
	=
	\frac{15-\nu}{8\mathcal L^4}
	-
	\frac{3}{\mathcal L^3\mathcal G}.
	\label{eq:averaged-1pn-hamiltonian}
\end{equation}
The conservative normal-form Hamiltonian through 1PN order is therefore
\begin{equation}
	\overline H(\mathcal L,\mathcal G)
	=
	-\frac{1}{2\mathcal L^2}
	+
	\varepsilon
	\left(
	\frac{15-\nu}{8\mathcal L^4}
	-
	\frac{3}{\mathcal L^3\mathcal G}
	\right)
	+
	\mathcal O(\varepsilon^2).
	\label{eq:normal-form-hamiltonian}
\end{equation}
When taking action derivatives, \(\mathcal L\) and \(\mathcal G\) are
treated as independent, while \(e\) is understood as the action-dependent
function defined in Eq.~\eqref{eq:kepler-definitions}. Final expressions may
subsequently be reparametrized by \((\mathcal L,e)\) using
\(\mathcal G=\mathcal L\sqrt{1-e^2}\).

The two normal-form frequencies are
\begin{subequations}
	\begin{align}
		&\omega_\ell
		=
		\frac{\partial\overline H}{\partial\mathcal L}=
		\frac{1}{\mathcal L^3}
		+
		\varepsilon
		\left[
		-\frac{15-\nu}{2\mathcal L^5}
		+
		\frac{9}{\mathcal L^4\mathcal G}
		\right]
		+
		\mathcal O(\varepsilon^2),
		\label{eq:radial-frequency}
		\\
		&\omega_g
		=
		\frac{\partial\overline H}{\partial\mathcal G}
		=
		\frac{3\varepsilon}
		{\mathcal L^3\mathcal G^2}
		+
		\mathcal O(\varepsilon^2).
		\label{eq:apsidal-frequency}
	\end{align}
\end{subequations}
Consequently, the radial period and the apsidal advance accumulated during
one radial cycle are
\begin{align}
	&T_r
	=
	\frac{2\pi}{\omega_\ell}
	=
	2\pi\mathcal L^3
	\left[
	1+
	\varepsilon
	\left(
	\frac{15-\nu}{2\mathcal L^2}
	-
	\frac{9}{\mathcal L\mathcal G}
	\right)
	\right]
	+
	\mathcal O(\varepsilon^2),
	\label{eq:radial-period-1pn}
	\\
	&\Delta g_{\rm cons}
	=
	\omega_g T_r
	=
	\frac{6\pi\varepsilon}{\mathcal G^2}
	+
	\mathcal O(\varepsilon^2).
	\label{eq:periastron-advance}
\end{align}
The usual periastron factor \cite{AIHPA_1985__43_1_107_0,Memmesheimer:2004cv},
equal to the total azimuthal phase accumulated per radial cycle divided by
$2\pi$, is 
\begin{equation}
	K
	=
	1+\frac{\omega_g}{\omega_\ell}
	=
	1+\frac{3\varepsilon}{\mathcal G^2}
	+
	\mathcal O(\varepsilon^2).
	\label{eq:periastron-factor}
\end{equation}

\subsection{Periodic Lie generator and ADM map}
\label{subsec:periodic-lie-generator}

The normal-form Hamiltonian is related to the physical ADM Hamiltonian by a
near-identity canonical transformation
\cite{Deprit_1969,Damour:2015isa}. We construct it explicitly with a
first-order Lie transform and use the Delaunay Poisson bracket
\begin{equation}
	\{A,B\}_{\rm D}
	=
	\frac{\partial A}{\partial\ell}
	\frac{\partial B}{\partial\mathcal L}
	-
	\frac{\partial A}{\partial\mathcal L}
	\frac{\partial B}{\partial\ell}
	+
	\frac{\partial A}{\partial g}
	\frac{\partial B}{\partial\mathcal G}
	-
	\frac{\partial A}{\partial\mathcal G}
	\frac{\partial B}{\partial g}.
	\label{eq:delaunay-poisson-bracket}
\end{equation}
With our sign convention, the first-order Lie generator \(W_1\) satisfies
\begin{equation}
	H_{\rm 1PN}
	+
	\{H_{\rm N},W_1\}_{\rm D}
	=
	\overline H_{\rm 1PN}.
	\label{eq:homological-equation}
\end{equation}
Because \(H_{\rm N}=-1/(2\mathcal L^2)\), this becomes
\begin{equation}
	\frac{\partial W_1}{\partial\ell}
	=
	\mathcal L^3
	\left(
	H_{\rm 1PN}
	-
	\overline H_{\rm 1PN}
	\right).
	\label{eq:homological-equation-explicit}
\end{equation}

Choosing the action-dependent integration constant so that \(W_1\) has
vanishing mean-anomaly average gives
\begin{align}
	W_1(u;\mathcal L,\mathcal G)
	&=
	\frac{e\sin u}{\mathcal L}
	\left(
	\frac{4-\nu}{2}
	-
	\frac{3}{s}
	+
	\frac{\nu}{2\rho}
	\right)
	\nonumber\\
	&\quad
	-
	\frac{3}{\mathcal L s}
	\bigl[v(u,e)-u\bigr].
	\label{eq:closed-form-w1}
\end{align}
This generator is single-valued over the radial cycle and obeys
\begin{align}
	&W_1(u+2\pi)=W_1(u),
	\qquad
	\left\langle W_1\right\rangle_\ell=0,
	\\
	&W_1(0)=W_1(2\pi)=0.
	\label{eq:w1-normalization}
\end{align}

The physical ADM map through 1PN order is then compactly specified by
\begin{equation}
	Z^a(\bm Y)
	=
	Z_{\rm N}^a(\bm Y)
	+
	\varepsilon
	\left\{
	Z_{\rm N}^a(\bm Y),W_1(\bm Y)
	\right\}_{\rm D}
	+
	\mathcal O(\varepsilon^2).
	\label{eq:adm-lie-map}
\end{equation}
Eq.~\eqref{eq:adm-lie-map}, together with
Eq.~\eqref{eq:closed-form-w1}, determines all four components of the 1PN
ADM embedding. Since \(W_1\) is independent of \(g\), the azimuthal action
is unchanged,
\begin{equation}
	p_\phi=\mathcal G+\mathcal O(\varepsilon^2).
	\label{eq:angular-momentum-map}
\end{equation}
Substitution of Eq.~\eqref{eq:adm-lie-map} into the ADM Hamiltonian reproduces
Eq.~\eqref{eq:normal-form-hamiltonian}, providing a direct check of the sign
and normalization of the canonical transformation.

\subsection{Time parametrization and Jacobi transport}
\label{subsec:time-jacobi-transport}

For an orbit beginning at periastron, with \(\ell_0=0\), the physical time
corresponding to the eccentric anomaly \(u\) follows from
\(\ell(t)=\omega_\ell t\) and Kepler's equation:
\begin{equation}
	t(u)
	=
	\frac{u-e\sin u}{\omega_\ell},
	\qquad
	\frac{\diff t}{\diff u}
	=
	\frac{\rho(u)}{\omega_\ell}.
	\label{eq:time-eccentric-anomaly}
\end{equation}
Both expressions are expanded through 1PN order before the
radiation-reaction integrand is formed. In particular,
\(t(2\pi)=T_r\), with \(T_r\) given by
Eq.~\eqref{eq:radial-period-1pn}.

The frequency Jacobian introduced in
Eq.~\eqref{eq:frequency-jacobian} is, through 1PN order,
\begin{widetext}
\begin{equation}
	D_{\bm J}\bm\omega
	=
	\begin{pmatrix}
		-\dfrac{3}{\mathcal L^4}
		+
		\varepsilon
		\left[
		\dfrac{5(15-\nu)}{2\mathcal L^6}
		-
		\dfrac{36}{\mathcal L^5\mathcal G}
		\right]
		&
		-\dfrac{9\varepsilon}
		{\mathcal L^4\mathcal G^2}
		\\[2ex]
		-\dfrac{9\varepsilon}
		{\mathcal L^4\mathcal G^2}
		&
		-\dfrac{6\varepsilon}
		{\mathcal L^3\mathcal G^3}
	\end{pmatrix}
	+
	\mathcal O(\varepsilon^2).
	\label{eq:frequency-jacobian-explicit}
\end{equation}
\end{widetext}
The corresponding fixed-time normal-form Jacobi matrix is given by
Eq.~\eqref{eq:delaunay-jacobi}.

For convenience, we denote the Jacobian of the ADM embedding by
\begin{equation}
	\mathsf C^a{}_{A}(\bm Y)
	\equiv
	\frac{\partial Z^a(\bm Y)}{\partial Y^A}.
	\label{eq:canonical-map-jacobian}
\end{equation}

Using this abbreviation, the tangent map already defined in
Eq.~\eqref{eq:composed-adm-tangent-map} can be written as
\begin{equation} \label{eq:K_compact}
	\mathsf K^a{}_{A}(t;\bm Y_0)
	=
	\mathsf C^a{}_{B}
	\bigl(\bm Y(t;\bm Y_0)\bigr)
	\left[
	\mathsf M_{\bm Y}(t)
	\right]^B{}_{A},
\end{equation}
where 
$
\bm Y(t;\bm Y_0)
=
\bigl(
\ell_0+\omega_\ell t,\,
g_0+\omega_g t,\,
\mathcal L,\,
\mathcal G
\bigr)
$.
In evaluating Eq.~\eqref{eq:K_compact}, the derivatives with respect to
the initial Delaunay variables are understood at fixed physical time
$t$. Only after the derivatives and the matrix composition have been
performed do we use $t=t(u)$ from
Eq.~\eqref{eq:time-eccentric-anomaly}. This order of operations avoids
introducing spurious contributions from the action dependence of
$t(u)$ into the fixed-time tangent map.

The remaining derivatives required to evaluate
$\mathsf C^a{}_{A}$ follow from Kepler's equation and the definition of
the true anomaly:
\begin{equation}
	\left.
	\frac{\partial u}{\partial e}
	\right|_\ell
	=
	\frac{\sin u}{\rho},
	\qquad
	\frac{\partial v}{\partial u}
	=
	\frac{s}{\rho},
	\qquad
	\left.
	\frac{\partial v}{\partial e}
	\right|_u
	=
	\frac{\sin u}{s\rho}.
	\label{eq:kepler-implicit-derivatives}
\end{equation}
Given the definition of the eccentricity parameter,
Eq.~\eqref{eq:ecc_param}, its required derivatives are
\begin{equation}
	\left.
	\frac{\partial e}{\partial\mathcal L}
	\right|_{\mathcal G}
	=
	\frac{s^2}{e\mathcal L},
	\qquad
	\left.
	\frac{\partial e}{\partial\mathcal G}
	\right|_{\mathcal L}
	=
	-\frac{s}{e\mathcal L}.
	\label{eq:eccentricity-action-derivatives}
\end{equation}
For later use, we now give explicitly the two rows of the tangent map
that enter the force pullback. We write
\begin{equation}
	\mathsf K^a{}_{A}(u)
	=
	\mathsf K^a{}_{{\rm N},A}(u)
	+
	\varepsilon\,
	\mathsf K^a{}_{{\rm 1PN},A}(u)
	+
	\mathcal O(\varepsilon^2),
	\label{eq:tangent-map-pn-explicit}
\end{equation}
with $A=(\ell,g,\mathcal L,\mathcal G)$ and
$\ell=\ell(u)=u-e\sin u$. At Newtonian order we obtain
\begin{widetext}
	\begin{subequations}
		\label{eq:newtonian-tangent-map-explicit}
		\begin{align}
			&\mathsf K^r{}_{{\rm N},A}
			=
			\Bigg(
			\frac{\mathcal L^2 e\sin u}{\rho},
			\,0,\,
			\mathcal L
			\left[
			2\rho
			+
			\frac{s^2(e-\cos u)}{e\rho}
			-
			\frac{3e\sin u\,\ell}{\rho}
			\right],
			\,
			-\frac{\mathcal L s(e-\cos u)}{e\rho}
			\Bigg),
			\\
			&\mathsf K^\phi{}_{{\rm N},A}
			=
			\Bigg(
			\frac{s}{\rho^2},
			\,1,\,
			\frac{s}{\mathcal L\rho^2}
			\left[
			\frac{\sin u}{e}
			\left(
			2-e\cos u-e^2
			\right)
			-
			3\ell
			\right],
			\,
			-\frac{\sin u}{e\mathcal L\rho^2}
			\left(
			2-e\cos u-e^2
			\right)
			\Bigg).
		\end{align}
	\end{subequations}
\end{widetext}

To write the relative-1PN part compactly, let
$r_{\rm 1PN}$ and $\phi_{\rm 1PN}$ denote the coefficients of
$\varepsilon$ in the ADM map \eqref{eq:adm-lie-map}, explicitly
\begin{widetext}
	\begin{subequations}
		\label{eq:radial-azimuthal-map-1pn}
		\begin{align}
			&r_{\rm 1PN}
			=
			-\frac{\nu}{2\rho}
			\left(
			3\rho^2-5\rho+2s^2
			\right)
			-\frac{6e^2\rho}{s(1+s)}-\frac{8s+5}{1+s}
			+\frac{s^2(4s+1)}{\rho(1+s)},
			\\
			&\phi_{\rm 1PN}
			=
			\frac{1}{\mathcal L^2}
			\Bigg\{
			\frac{3\bigl[v(u,e)-u\bigr]}{s^2}
			+
			\frac{e\sin u}{\rho^2s^2}
			\left[
			\nu s^3
			+
			3\rho^2
			+
			\frac{s\left[3\rho-s^2(4s+1)\right]}{1+s}
			\right]
			\Bigg\}.
		\end{align}
	\end{subequations}
\end{widetext}
It is also convenient to introduce the derivative at fixed mean
anomaly,
\begin{equation}
	\partial_e^{(\ell)}
	\equiv
	\left.
	\frac{\partial}{\partial e}
	\right|_\ell
	=
	\left.
	\frac{\partial}{\partial e}
	\right|_u
	+
	\frac{\sin u}{\rho}
	\frac{\partial}{\partial u}.
	\label{eq:fixed-ell-e-derivative}
\end{equation}
When this operator acts on $v(u,e)$, the fixed-$u$ derivative is the one
given in Eq.~\eqref{eq:kepler-implicit-derivatives}.

The relative-1PN tangent-map coefficients can then be written in the
compact explicit form
\begin{subequations}
	\label{eq:1pn-tangent-map-explicit}
	\begin{align}
		\mathsf K^r{}_{{\rm 1PN},\ell}
		&=
		\frac{1}{\rho}
		\frac{\partial r_{\rm 1PN}}{\partial u},
		\qquad
		\mathsf K^r{}_{{\rm 1PN},g}=0,
		\\
		\mathsf K^r{}_{{\rm 1PN},\mathcal L}
		&=
		\frac{s^2}{e\mathcal L}
		\partial_e^{(\ell)} r_{\rm 1PN}
		-
		\frac{3\ell}{\mathcal L}
		\mathsf K^r{}_{{\rm 1PN},\ell}
		\\
		&\qquad+
		\frac{\ell e\sin u}{\mathcal L\rho}
		\left(
		15-\nu-\frac{9}{s}
		\right),
		\\
		\mathsf K^r{}_{{\rm 1PN},\mathcal G}
		&=
		-\frac{s}{e\mathcal L}
		\partial_e^{(\ell)} r_{\rm 1PN}
		-
		\frac{9\ell e\sin u}
		{\mathcal L s^2\rho},
		\label{eq:1pn-tangent-r}
		\\  \mathsf K^\phi{}_{{\rm 1PN},\ell}&=
		\frac{1}{\rho}
		\frac{\partial\phi_{\rm 1PN}}{\partial u}, \qquad
		\mathsf K^\phi{}_{{\rm 1PN},g}=0,
		\\
		\mathsf K^\phi{}_{{\rm 1PN},\mathcal L}
		&=
		-\frac{2}{\mathcal L}\phi_{\rm 1PN}
		+
		\frac{s^2}{e\mathcal L}
		\partial_e^{(\ell)}\phi_{\rm 1PN}
		-
		\frac{3\ell}{\mathcal L}
		\mathsf K^\phi{}_{{\rm 1PN},\ell}
		\nonumber\\
		&\qquad
		+
		\frac{\ell}{\mathcal L^3}
		\left[
		\frac{s(15-\nu)-9}{\rho^2}
		-
		\frac{9}{s^2}
		\right],
		\\
		\mathsf K^\phi{}_{{\rm 1PN},\mathcal G}
		&=
		-\frac{s}{e\mathcal L}
		\partial_e^{(\ell)}\phi_{\rm 1PN}
		-
		\frac{\ell}{\mathcal L^3}
		\left(
		\frac{9}{s\rho^2}
		+
		\frac{6}{s^3}
		\right).
		\label{eq:1pn-tangent-phi}
	\end{align}
\end{subequations}

Eqs.~\eqref{eq:newtonian-tangent-map-explicit} and
\eqref{eq:1pn-tangent-map-explicit}, together with
Eq.~\eqref{eq:time-eccentric-anomaly}, provide the conservative
tangent-map input required by the relative-1PN one-cycle pullback.

\section{One-cycle Magnusian through relative 1PN order}
\label{sec:one-cycle-result}

We now combine the conservative ingredients constructed in
Sec.~\ref{sec:conservative-normal-form} with the 2.5PN and 3.5PN
radiation-reaction forces introduced in Sec.~\ref{subsec:pn-input}.

\subsection{Expansion of the pulled-back force}
\label{subsec:nlo-pullback-expansion}

After changing the integration variable from physical time to eccentric
anomaly, Eq.~\eqref{eq:one-cycle-magnusian-integral} becomes
\begin{align}
	\chi_A^{(1)}
	&=
	\int_0^{2\pi}\!\diff u\,
	\frac{\diff t}{\diff u}
	\mathcal F_r[\bm z^{(0)}(u)]
	\,\mathsf K^r{}_{A}(u)
	\nonumber\\
	&\quad
	+
	\int_0^{2\pi}\!\diff u\,
	\frac{\diff t}{\diff u}
	\mathcal F_\phi[\bm z^{(0)}(u)]
	\,\mathsf K^\phi{}_{A}(u),
	\label{eq:one-cycle-u-integral}
\end{align}
where \(A\in\{\ell,g,\mathcal L,\mathcal G\}\). All quantities in the
integrand are understood as functions of the initial Delaunay data
\(\bm Y_0=(0,0,\mathcal L,\mathcal G)\).

To organize the relative-1PN expansion, we write
\begin{subequations}
	\begin{align}
		&\bm z^{(0)}(u)
		=
		\bm z_{\rm N}(u)
		+
		\varepsilon\,\bm z_{\rm 1PN}(u)
		+
		\mathcal O(\varepsilon^2),
		\label{eq:orbit-pn-expansion}
		\\
		&\mathsf K^a{}_{A}(u)=
		\mathsf K^a{}_{{\rm N},A}(u)
		+
		\varepsilon\,
		\mathsf K^a{}_{{\rm 1PN},A}(u)
		+
		\mathcal O(\varepsilon^2),
		\label{eq:tangent-pn-expansion}
		\\
		&\frac{\diff t}{\diff u}
		=
		\mathcal T_{\rm N}(u)
		+
		\varepsilon\mathcal T_{\rm 1PN}(u)
		+
		\mathcal O(\varepsilon^2),
		\label{eq:measure-pn-expansion}
	\end{align}
\end{subequations}
where Eq.~\eqref{eq:time-eccentric-anomaly} gives
\begin{subequations}
	\begin{align}
		&\mathcal T_{\rm N}
		=
		\mathcal L^3\rho,
		\label{eq:newtonian-time-measure}
		\\
		&\mathcal T_{\rm 1PN}
		=
		\mathcal L^3\rho
		\left(
		\frac{15-\nu}{2\mathcal L^2}
		-
		\frac{9}{\mathcal L\mathcal G}
		\right),
		\label{eq:1pn-time-measure}
	\end{align}
\end{subequations}
while the relevant \(\mathsf K^a{}_{{\rm N},A}\) and
\(\mathsf K^a{}_{{\rm 1PN},A}\) components are given, respectively, in
Eqs.~\eqref{eq:newtonian-tangent-map-explicit} and
\eqref{eq:1pn-tangent-map-explicit}.
The Newtonian orbit $\bm z_{\rm N}$ is simply obtained by evaluating the
Kepler map in Eq.~\eqref{eq:newtonian-kepler-map} along the Newtonian
normal-form flow, and the relative-1PN correction
$\bm z_{\rm 1PN}$ is determined by
the Lie map in Eq.~\eqref{eq:adm-lie-map}.

We similarly expand the generalized polar force as
\begin{equation}
	\mathcal F_a
	=
	\lambda
	\left(
	\mathcal F_a^{\rm 2.5PN}
	+
	\varepsilon\mathcal F_a^{\rm 3.5PN}
	\right)
	+
	\mathcal O(\lambda\varepsilon^2),
	\quad
	a\in\{r,\phi\}.
	\label{eq:polar-force-pn-expansion}
\end{equation}
Its evaluation on the conservative orbit gives
\begin{align}
	\mathcal F_a[\bm z^{(0)}]
	&=
	\lambda
	\Bigg\{
	\mathcal F_a^{\rm 2.5PN}[\bm z_{\rm N}]
	\nonumber\\
	&\qquad
	+
	\varepsilon
	\left[
	\mathcal F_a^{\rm 3.5PN}[\bm z_{\rm N}]
	+
	z_{\rm 1PN}^b
	\frac{\partial\mathcal F_a^{\rm 2.5PN}}
	{\partial z^b}
	\bigg|_{\bm z_{\rm N}}
	\right]
	\Bigg\}
	\nonumber\\
	&\qquad+
	\mathcal O(\lambda\varepsilon^2),
	\label{eq:force-orbit-expansion}
\end{align}
where the index \(b\) runs over all four ADM variables.

Defining the eccentric-anomaly integrand by
\begin{equation}
	\chi_A^{(1)}
	\equiv
	\int_0^{2\pi}\!\diff u\,
	\mathcal I_A^{(1)}(u),
\end{equation}
we obtain
\begin{equation}
	\mathcal I_A^{(1)}
	=
	\mathcal I_{{\rm N},A}^{(1)}
	+
	\varepsilon\mathcal I_{{\rm 1PN},A}^{(1)}
	+
	\mathcal O(\varepsilon^2),
	\label{eq:integrand-pn-expansion}
\end{equation}
with
\begin{equation}
	\mathcal I_{{\rm N},A}^{(1)}
	=
	\lambda\,
	\mathcal T_{\rm N}
	\mathcal F_i^{\rm 2.5PN}
	\mathsf K^i{}_{{\rm N},A},
		\label{eq:lo-pullback-integrand}
\end{equation}
and
\begin{align}
	&\mathcal I_{{\rm 1PN},A}^{(1)}
	=
	\lambda\mathcal T_{\rm N}
	\mathcal F_i^{\rm 3.5PN}
	\mathsf K^i{}_{{\rm N},A}
	\nonumber\\
	&\quad+
	\lambda\mathcal T_{\rm N}
	\mathcal F_i^{\rm 2.5PN}
	\mathsf K^i{}_{{\rm 1PN},A}
	\nonumber\\
	&\quad+
	\lambda\mathcal T_{\rm N}
	z_{\rm 1PN}^b
	\frac{\partial\mathcal F_i^{\rm 2.5PN}}
	{\partial z^b}
	\mathsf K^i{}_{{\rm N},A}
	\nonumber\\
	&\quad+
	\lambda\mathcal T_{\rm 1PN}
	\mathcal F_i^{\rm 2.5PN}
	\mathsf K^i{}_{{\rm N},A}.
		\label{eq:nlo-pullback-integrand}
\end{align}

All forces and derivatives on the right-hand sides of
Eqs.~\eqref{eq:lo-pullback-integrand} and
\eqref{eq:nlo-pullback-integrand} are evaluated on the Newtonian orbit.
The four terms of Eq.~\eqref{eq:nlo-pullback-integrand} represent,
respectively, the direct 3.5PN force, the 1PN correction to the Jacobi
transport, the correction from evaluating the leading force on the 1PN
ADM orbit, and the 1PN correction to the time measure.

\subsection{Exact-in-eccentricity integration over one radial cycle}
\label{subsec:exact-radial-integration}

The individual terms in Eq.~\eqref{eq:nlo-pullback-integrand} contain the
continuous angles \(u\), \(\ell=u-e\sin u\), and \(v(u,e)\). These
nonperiodic terms cancel after the two halves of the radial cycle are
combined. We therefore introduce the paired integrand
\begin{align}
	\mathcal I_{{\rm 1PN},A}^{(1),{\rm pair}}(u)
	&\equiv
	\mathcal I_{{\rm 1PN},A}^{(1)}(u)
	+
	\mathcal I_{{\rm 1PN},A}^{(1)}(2\pi-u),
	\label{eq:paired-nlo-integrand}
\end{align}
with $0\leq u\leq\pi$.
The required reflection identities are
\begin{equation}
	\ell(2\pi-u)=2\pi-\ell(u),
	\qquad
	v(2\pi-u,e)=2\pi-v(u,e).
	\label{eq:angle-reflection-identities}
\end{equation}
After applying these identities, all explicit dependence on the unwrapped
quantities $\ell-\pi$ and $v-u$ cancels in the paired integrand
$\mathcal I^{\rm pair}_{{\rm 1PN},A}(u)$. It therefore becomes a rational function of \(\sin u\), \(\cos u\), \(e\), and
\(s=\sqrt{1-e^2}\).

We rationalize the remaining integral by introducing
\begin{align}
	&x=\tan\frac{u}{2},
	\quad
	\sin u=\frac{2x}{1+x^2},
	\quad
	\cos u=\frac{1-x^2}{1+x^2},
	\\&
	\diff u=\frac{2\,\diff x}{1+x^2}.
	\label{eq:half-angle-substitution}
\end{align}
In particular we have
\begin{equation}
	\rho(u)
	=
	\frac{(1-e)+(1+e)x^2}{1+x^2}.
	\label{eq:rho-half-angle}
\end{equation}
Each 1PN component then reduces to a finite linear combination of integrals
of the form
\begin{equation}
	\int_0^\infty\!\diff x\,
	\frac{x^{2k}}
	{\left[(1-e)+(1+e)x^2\right]^n}.
	\label{eq:half-angle-moment}
\end{equation}
For the four components ordered as
\((\ell,g,\mathcal L,\mathcal G)\), the denominator orders are
\begin{equation}
	(n_\ell,n_g,n_{\mathcal L},n_{\mathcal G})
	=
	(8,6,8,6),
\end{equation}
and the corresponding numerator degrees in \(x\) are $(14,10,14,10)$.
The required moments are evaluated exactly using
\begin{align}
	&\int_0^\infty\!\diff x\,
	\frac{x^{2k}}{(c_1+c_2 x^2)^n}
	\nonumber\\
	&\qquad
	=
	\frac{1}{2}
	c_1^{\,k+\frac12-n}
	c_2^{-k-\frac12}
	\mathrm B
	\left(
	k+\frac12,\,
	n-k-\frac12
	\right),
	\label{eq:beta-function-moment}
\end{align}
where \(\mathrm B\) denotes the Euler beta
function and, in our case, \(c_1=1-e\), \(c_2=1+e\). For bound eccentricity
\(0\leq e<1\), the moments required here satisfy \(c_1>0\), \(c_2>0\), and
\(n>k+1/2\). This gives all four one-cycle components in closed form without
an expansion in eccentricity.

\subsection{Closed-form result}
\label{subsec:closed-form-magnusian}

We now write down the one-cycle covector through relative 1PN order in the Delaunay
basis of Eq.~\eqref{eq:chi1_exp_eps}.

It is convenient to define
\begin{equation}
	\mathcal P_0(e)
	\equiv
	96+292e^2+37e^4.
	\label{eq:leading-eccentricity-polynomial}
\end{equation}
In these terms the leading coefficient is
\begin{equation}
\chi_{A,{\rm N}}^{(1)}
	=
	\lambda\left(
	-\frac{2\nu\pi\,\mathcal P_0}
	{15s^7\mathcal L^4},
	-\frac{8\nu\pi(8+7e^2)}
	{5s^4\mathcal L^4},
	\frac{2\nu\pi^2\,\mathcal P_0}
	{5s^7\mathcal L^5},
	0
	\right).
	\label{eq:lo-magnusian}
\end{equation}
Its reduction to the computation
of Ref.~\cite{Blanco:2026prj} will be discussed separately below.

The relative-1PN coefficient can be written compactly as
\begin{subequations}
	\label{eq:nlo-magnusian-components}
	\begin{align}
		\chi_{\ell,{\rm 1PN}}^{(1)}
		&=
		\lambda\frac{\nu\pi}
		{420s^9\mathcal L^6}
		\mathcal P_\ell(e,s,\nu),
		\label{eq:nlo-chi-ell}
		\\
		\chi_{g,{\rm 1PN}}^{(1)}
		&=
		\lambda\frac{\nu\pi}
		{210s^6\mathcal L^6}
		\mathcal P_g(e,s,\nu),
		\label{eq:nlo-chi-g}
		\\
		\chi_{\mathcal L,{\rm 1PN}}^{(1)}
		&=
	\lambda	\frac{\nu\pi^2}
		{420s^9\mathcal L^7}
		\mathcal P_{\mathcal L}(e,s,\nu),
		\label{eq:nlo-chi-L}
		\\
		\chi_{\mathcal G,{\rm 1PN}}^{(1)}
		&=
	\lambda	\frac{6\nu\pi^2}
		{5s^9\mathcal L^7}
		\mathcal P_{\mathcal G}(e).
		\label{eq:nlo-chi-G}
	\end{align}
\end{subequations}
The eccentricity polynomials appearing here are
\begin{subequations}
	\begin{align}
		\mathcal P_\ell
		={}&
		16\left(
		-11941+10164s+756\nu
		\right)
		\nonumber\\
		&
		+
		8e^2\left(
		-15795+8988s+10472\nu
		\right)
		\nonumber\\
		&
		+
		42e^4\left(
		891-484s+1400\nu
		\right)
		\nonumber\\
		&
		+
		e^6\left(
		-2393+3108\nu
		\right),
		\label{eq:polynomial-Pell}
		\\[1ex]
		\mathcal P_g
		={}&
		8\left(
		-3541+1764s+756\nu
		\right)
		\nonumber\\
		&
		+
		8e^2\left(
		467-1764s+2786\nu
		\right)
		\nonumber\\
		&
		+
		e^4\left(
		-13103+4564\nu
		\right),
		\label{eq:polynomial-Pg}
		\\[1ex]
		\mathcal P_{\mathcal L}
		={}&
		48\left(
		10261-8148s-644\nu
		\right)
		\nonumber\\
		&
		+
		8e^2\left(
		26805-9324s-30044\nu
		\right)
		\nonumber\\
		&
		+
		42e^4\left(
		2427+888s-4540\nu
		\right)
		\nonumber\\
		&
		+
		e^6\left(
		38259-11396\nu
		\right),
		\label{eq:polynomial-PL}
		\\[1ex]
		\mathcal P_{\mathcal G}
		={}&
		160+284e^2-19e^4.
		\label{eq:polynomial-PG}
	\end{align}
\end{subequations}
Finally, the action changes and interaction-picture angle kicks generated by
the first Magnusian are obtained from
Eq.~\eqref{eq:delaunay-kick-from-chi}. In the action sector,~\footnote{We recall that the coefficient covector $\chi_A^{(1)}$ defined in Eq.~\eqref{eq:covector_def} is linear in $\lambda$.}
\begin{align}
	\Delta\mathcal L_{\rm RR}
	&=
	\chi_{\ell,{\rm N}}^{(1)}
	+
	\varepsilon\chi_{\ell,{\rm 1PN}}^{(1)}
	+
	\mathcal O(\lambda\varepsilon^2,\lambda^2),
	\\
	\Delta\mathcal G_{\rm RR}
	&=
	\chi_{g,{\rm N}}^{(1)}
	+
	\varepsilon\chi_{g,{\rm 1PN}}^{(1)}
	+
	\mathcal O(\lambda\varepsilon^2,\lambda^2).
	\label{eq:physical-action-losses}
\end{align}
The remaining two components determine the corresponding interaction-picture angle kicks,
\begin{align}
	\Delta\ell_{\rm RR}
	&=
	-
	\chi_{\mathcal L,\rm N}^{(1)}
	-
	\varepsilon\chi_{\mathcal L,\rm 1PN}^{(1)}
	+
	\mathcal O(\lambda\varepsilon^2,\lambda^2),
	\\
	\Delta g_{\rm RR}
	&=
	-
	\chi_{\mathcal G,\rm N}^{(1)}
	-
	\varepsilon\chi_{\mathcal G,\rm 1PN}^{(1)}
	+
	\mathcal O(\lambda\varepsilon^2,\lambda^2).
\end{align}
These quantities belong to the interaction-picture one-cycle construction
and should be distinguished from the physical timing and apsidal shifts on
the perturbed return section, as specifically discussed below in
Sec.~\ref{subsec:return-section}.

We recall that Eqs.~\eqref{eq:nlo-magnusian-components} have been obtained using the
ADM representative of the 3.5PN radiation-reaction force. In
Sec.~\ref{sec:gauge} we will repeat the computation for the complete eight-parameter
Iyer--Will family and show that every gauge-dependent contribution integrates
to zero.

\section{Physical balance and structural checks}
\label{sec:physical-checks}

The closed-form result of Sec.~\ref{sec:one-cycle-result} admits several
independent checks. Energy and angular-momentum balance test the two
components \(\chi_\ell^{(1)}\) and \(\chi_g^{(1)}\) that determine the action
changes, while the geometry of the normal-form shear relates the action and
angle sectors and thereby constrains the complete four-component covector.
We then examine the circular limit and compare the induced evolution of the
radial quasi-Keplerian elements with known flux results for \(0<e<1\), without
an eccentricity expansion.

\subsection{Energy and angular-momentum balance}
\label{subsec:energy-angular-balance}

The direct mechanical work and torque produced by the generalized polar
forces satisfy
\begin{equation}
	\frac{\diff H_{\rm cons}}{\diff t}
	=
	\dot r\,\mathcal F_r
	+
	\dot\phi\,\mathcal F_\phi,
	\qquad
	\frac{\diff p_\phi}{\diff t}
	=
	\mathcal F_\phi.
	\label{eq:direct-work-torque}
\end{equation}
Their changes over one conservative radial cycle are therefore
\begin{align}
	\Delta E_{\rm dir}
	&=
	\int_0^{T_r}\!\diff t\,
	\left(
	\dot r\,\mathcal F_r
	+
	\dot\phi\,\mathcal F_\phi
	\right),
	\label{eq:direct-energy-change}
	\\
	\Delta J_{\rm dir}
	&=
	\int_0^{T_r}\!\diff t\,
	\mathcal F_\phi.
	\label{eq:direct-angular-momentum-change}
\end{align}

The same changes follow directly from the components \(\chi_\ell^{(1)}\) and
\(\chi_g^{(1)}\). Since the conservative Hamiltonian in normal-form variables
depends only on the actions, we have
\begin{equation}
	\Delta E_\chi
	=
	\omega_\ell\Delta\mathcal L
	+
	\omega_g\Delta\mathcal G
	=
	\omega_\ell\chi_\ell^{(1)}
	+
	\omega_g\chi_g^{(1)}
	+
	\mathcal O(\lambda^2),
	\label{eq:energy-change-from-magnusian}
\end{equation}
while \(p_\phi=\mathcal G+\mathcal O(\varepsilon^2)\) implies
\begin{equation}
	\Delta J_\chi
	=
	\Delta\mathcal G
	=
	\chi_g^{(1)}
	+
	\mathcal O(\lambda^2).
	\label{eq:angular-change-from-magnusian}
\end{equation}

To display the relative-1PN content explicitly, we expand the normal-form
frequencies as
\begin{subequations}
	\begin{align}
		\omega_\ell
		&=
		\omega_{\ell,{\rm N}}
		+
		\varepsilon\omega_{\ell,{\rm 1PN}}
		+
		\mathcal O(\varepsilon^2),
		\\
		\omega_g
		&=
		\varepsilon\omega_{g,{\rm 1PN}}
		+
		\mathcal O(\varepsilon^2),
	\end{align}
\end{subequations}
where the periastron-precession frequency vanishes at Newtonian order.
Writing
\begin{equation}
	\Delta E_\chi
	=
	\Delta E_{{\rm N},\chi}
	+
	\varepsilon\Delta E_{{\rm 1PN},\chi}
	+
	\mathcal O(\lambda\varepsilon^2,\lambda^2),
\end{equation}
and analogously for \(\Delta J_\chi\), the corresponding coefficients read
\begin{subequations}
	\label{eq:energy-balance}
	\begin{align}
		\Delta E_{{\rm N},\chi}
		&=
		\omega_{\ell,{\rm N}}
		\chi_{\ell,{\rm N}}^{(1)},
		\label{eq:lo-energy-balance}
		\\
		\Delta E_{{\rm 1PN},\chi}
		&=
		\omega_{\ell,{\rm N}}
		\chi_{\ell,{\rm 1PN}}^{(1)}
		+
		\omega_{\ell,{\rm 1PN}}
		\chi_{\ell,{\rm N}}^{(1)}
		\nonumber\\&\qquad+
		\omega_{g,{\rm 1PN}}
		\chi_{g,{\rm N}}^{(1)},
		\label{eq:nlo-energy-balance}
	\end{align}
\end{subequations}
and
\begin{equation}
	\Delta J_{{\rm N},\chi}
	=
	\chi_{g,{\rm N}}^{(1)},
	\qquad
	\Delta J_{{\rm 1PN},\chi}
	=
	\chi_{g,{\rm 1PN}}^{(1)}.
	\label{eq:angular-momentum-balance}
\end{equation}

Substituting the explicit Magnusian integrands into
Eqs.~\eqref{eq:energy-balance}-\eqref{eq:angular-momentum-balance}
shows that, through relative 1PN order, the resulting integrands for the
energy and angular-momentum changes coincide with the direct mechanical
work and torque integrands in
Eqs.~\eqref{eq:direct-energy-change} and
\eqref{eq:direct-angular-momentum-change}.
Consequently,
\begin{equation}
	\left(
	\Delta E_\chi,
	\Delta J_\chi
	\right)
	=
	\left(
	\Delta E_{\rm dir},
	\Delta J_{\rm dir}
	\right)
	+
	\mathcal O(\lambda\varepsilon^2,\lambda^2),
	\label{eq:integrated-work-torque-balance}
\end{equation}
for arbitrary bound eccentricity. Since the equality holds before integration,
this provides a direct check of both the force conversion and the Jacobi
transport entering the first two components of $\bm\chi^{(1)}$.

\subsection{Midpoint-shear identity}
\label{subsec:midpoint-shear}

The two remaining components of the Magnusian are constrained by the shear
structure of the normal-form Jacobi propagator. We divide the coefficient
covector into its angle-indexed and action-indexed blocks,
\begin{equation}
	\bm\chi_{\bm\theta}^{(1)}
	\equiv
	\begin{pmatrix}
		\chi_\ell^{(1)}\\
		\chi_g^{(1)}
	\end{pmatrix},
	\qquad
	\bm\chi_{\bm J}^{(1)}
	\equiv
	\begin{pmatrix}
		\chi_{\mathcal L}^{(1)}\\
		\chi_{\mathcal G}^{(1)}
	\end{pmatrix}.
	\label{eq:magnusian-blocks}
\end{equation}
The angle-indexed block gives the physical action changes, whereas the
negative of the action-indexed block gives the interaction-picture angle
kick.

The origin of the relation between the two blocks can be seen directly from
the shear structure of the normal-form Jacobi propagator. Considering
Eq.~\eqref{eq:delaunay-jacobi} together with Eq.~\eqref{eq:K_compact}, the angle and action columns of the
ADM tangent map have the schematic form
\begin{equation}
	\mathsf K_{\bm\theta}
	=
	\mathsf C_{\bm\theta},
	\qquad
	\mathsf K_{\bm J}
	=
	\mathsf C_{\bm J}
	+
	t\,\mathsf C_{\bm\theta}
	D_{\bm J}\bm\omega.
\end{equation}
Thus, an initial variation of the actions produces not only a direct
variation of the orbit, but also an accumulated variation of the angles
through the conservative frequencies.

After pairing the two halves of the radial cycle, the direct contribution
associated with $\mathsf C_{\bm J}$ has vanishing one-cycle integral, whereas
the surviving shear contribution is determined by the same periodic
pullback that enters the angle block. Denoting this two-component
physical-force pullback by $\bm f(t)$, one may therefore write
\begin{align}
	\bm\chi_{\bm\theta}^{(1)}
	&=
	\int_0^{T_r}\!\diff t\,\bm f(t),
	\\
	\bm\chi_{\bm J}^{(1)}
	&=
	\left(
	D_{\bm J}\bm\omega
	\right)
	\int_0^{T_r}\!\diff t\,
	t\,\bm f(t),
	\label{eq:midpoint-shear-integrals}
\end{align}
where we have used the symmetry of
$D_{\bm J}\bm\omega$, which is the Hessian of the normal-form Hamiltonian.

Reflection about the midpoint of the radial cycle gives
\begin{equation}
	\bm f(T_r-t)=\bm f(t).
\end{equation}
Consequently,
\begin{align}
	\int_0^{T_r}\!\diff t\,
	t\,\bm f(t)
	&=
	\int_0^{T_r}\!\diff t\,
	(T_r-t)\,\bm f(t)
	\nonumber\\
	&=
	T_r
	\int_0^{T_r}\!\diff t\,\bm f(t)
	-
	\int_0^{T_r}\!\diff t\,
	t\,\bm f(t),
\end{align}
and hence
\begin{equation}
	\int_0^{T_r}\!\diff t\,
	t\,\bm f(t)
	=
	\frac{T_r}{2}
	\int_0^{T_r}\!\diff t\,
	\bm f(t).
	\label{eq:midpoint-reflection-moment}
\end{equation}
Substitution into Eq.~\eqref{eq:midpoint-shear-integrals} then yields
\begin{equation}
	\bm\chi_{\bm J}^{(1)}
	=
	\frac{T_r}{2}
	\left(
	D_{\bm J}\bm\omega
	\right)
	\bm\chi_{\bm\theta}^{(1)}.
	\label{eq:midpoint-shear-identity}
\end{equation}

Writing
\begin{subequations}
	\begin{align}
		T_r
		&=
		T_{r,{\rm N}}
		+
		\varepsilon T_{r,{\rm 1PN}}
		+
		\mathcal O(\varepsilon^2),
		\\
		D_{\bm J}\bm\omega
		&=
		\mathsf D_{\rm N}
		+
		\varepsilon\mathsf D_{\rm 1PN}
		+
		\mathcal O(\varepsilon^2),
	\end{align}
\end{subequations}
the midpoint-shear identity gives
\begin{subequations}
	\begin{align}
		\bm\chi_{\bm J,{\rm N}}^{(1)}
		&=
		\frac{T_{r,{\rm N}}}{2}
		\mathsf D_{\rm N}
		\bm\chi_{\bm\theta,{\rm N}}^{(1)},
		\label{eq:lo-midpoint-shear}
		\\
		\bm\chi_{\bm J,{\rm 1PN}}^{(1)}
		&=
		\frac{T_{r,{\rm N}}}{2}
		\mathsf D_{\rm N}
		\bm\chi_{\bm\theta,{\rm 1PN}}^{(1)}
		+
		\frac{T_{r,{\rm N}}}{2}
		\mathsf D_{\rm 1PN}
		\bm\chi_{\bm\theta,{\rm N}}^{(1)}
		\nonumber\\
		&\quad
		+
		\frac{T_{r,{\rm 1PN}}}{2}
		\mathsf D_{\rm N}
		\bm\chi_{\bm\theta,{\rm N}}^{(1)}.
		\label{eq:nlo-midpoint-shear}
	\end{align}
\end{subequations}
Substitution of the four closed-form components obtained in
Sec.~\ref{subsec:closed-form-magnusian} verifies both identities exactly.
Conversely, Eq.~\eqref{eq:midpoint-shear-identity} shows that the complete
first-order one-cycle Magnusian can be reconstructed from
\(\bm\chi_{\bm\theta}^{(1)}\), the two components that determine the action changes.

\subsection{Projection onto the periastron return section}
\label{subsec:return-section}

The first Magnusian constructed above describes the dissipative perturbation
over the fixed conservative interval $[0,T_r]$. This fixed-time endpoint
should be distinguished from the true return of the perturbed orbit to the
next periastron section. We now make this distinction explicit.

With the notation of the previous subsection, the interaction-picture kick generated by the first Magnusian at the
initial point is
\begin{equation}
	\delta\bm J
	=
	\bm\chi_{\bm\theta}^{(1)},
	\qquad
	\delta\bm\theta_{\rm raw}
	=
	-\bm\chi_{\bm J}^{(1)}.
\end{equation}
Propagating this perturbation to the fixed conservative endpoint $T_r$
with the normal-form Jacobi flow gives
\begin{equation}
	\delta\bm\theta_{\rm fix}
	=
	-\bm\chi_{\bm J}^{(1)}
	+
	T_r
	\left(D_{\bm J}\bm\omega\right)
	\bm\chi_{\bm\theta}^{(1)}.
	\label{eq:fixed-time-angle-shift}
\end{equation}
Using the midpoint-shear identity
\eqref{eq:midpoint-shear-identity}, this simplifies to
\begin{equation}
	\delta\bm\theta_{\rm fix}
	=
	\bm\chi_{\bm J}^{(1)}.
	\label{eq:fixed-endpoint-angle-shift}
\end{equation}

The true return time is obtained by requiring that the perturbed endpoint
lie again on the radial section. To first order,
\begin{equation}
	\delta\ell_{\rm fix}
	+
	\omega_\ell\,\delta T_{\rm ret}
	=
	0.
\end{equation}
Since Eq.~\eqref{eq:fixed-endpoint-angle-shift} gives
$\delta\ell_{\rm fix}=\chi_{\mathcal L}^{(1)}$, we obtain
\begin{equation}
	\delta T_{\rm ret}
	=
	-\frac{\chi_{\mathcal L}^{(1)}}{\omega_\ell}
	=
	\frac{\Delta\ell_{\rm RR}}{\omega_\ell},
	\label{eq:return-time-correction}
\end{equation}
where
$\Delta\ell_{\rm RR}\equiv-\chi_{\mathcal L}^{(1)}$
denotes the raw interaction-picture mean-anomaly kick.

The projection also modifies the apsidal angle. Its dissipative shift on
the return section is
\begin{align}
	\delta g_{\rm ret}
	&=
	\delta g_{\rm fix}
	+
	\omega_g\,\delta T_{\rm ret}
	\nonumber\\
	&=
	\chi_{\mathcal G}^{(1)}
	-
	\frac{\omega_g}{\omega_\ell}
	\chi_{\mathcal L}^{(1)}.
	\label{eq:return-apsidal-shift}
\end{align}
Because $\omega_g=\mathcal O(\varepsilon)$, the return-time contribution
to the apsidal shift first enters at relative 1PN order. Substitution of
the closed-form Magnusian gives
\begin{equation}
	\delta g_{\rm ret}
	=
	\lambda\varepsilon\,
	\frac{48\pi^2\nu}{5\mathcal G^7}
	\left(8+7e^2\right)
	+
	\mathcal O(\lambda\varepsilon^2,\lambda^2).
	\label{eq:return-apsidal-shift-explicit}
\end{equation}

The endpoint projection does not modify the action changes at $\mathcal O(\lambda)$, since the additional dissipative evolution over $\delta T_{\rm ret}=\mathcal O(\lambda)$ contributes only at $\mathcal O(\lambda^2)$. By contrast, it affects the return time already at $\mathcal O(\lambda)$ and induces an apsidal correction at $\mathcal O(\lambda\varepsilon)$, both of which lie within the accuracy considered here.

\subsection{Circular limit}
\label{subsec:circular-limit-check}

The action sector has a well-defined circular limit, which provides a direct
comparison with the standard 1PN energy and angular-momentum fluxes.
In this limit
\(\mathcal G=\mathcal L\), and the azimuthal frequency is
\begin{equation}
	\omega_\phi
	=
	\omega_\ell+\omega_g.
	\label{eq:circular-azimuthal-frequency}
\end{equation}
Introducing the usual invariant frequency parameter
\begin{equation}
	x
	\equiv
	\omega_\phi^{2/3},
\end{equation}
the orbit-averaged reduced mechanical loss rates through relative 1PN order
are, in our notation,\footnote{Division of the usual total fluxes by
	the reduced mass leaves one power of \(\nu\).}
\begin{align}
	&\langle\dot E\rangle_{\rm circ}
	=
	-\lambda\frac{32\nu}{5}x^5
	\left[
	1+
	\varepsilon x
	\left(
	-\frac{1247}{336}
	-\frac{35}{12}\nu
	\right)
	\right]
	\nonumber\\
	&\quad
	+
	\mathcal O(\lambda\varepsilon^2),
	\label{eq:circular-energy-flux}
	\\
	&\langle\dot J\rangle_{\rm circ}
	=
	-\lambda\frac{32\nu}{5}x^{7/2}
	\left[
	1+
	\varepsilon x
	\left(
	-\frac{1247}{336}
	-\frac{35}{12}\nu
	\right)
	\right]
	\nonumber\\
	&\quad
	+
	\mathcal O(\lambda\varepsilon^2),
	\label{eq:circular-angular-flux}
\end{align}
as can be found, for example, in
Ref.~\cite{Blanchet:2013haa}.

Multiplying Eqs.~\eqref{eq:circular-energy-flux} and
\eqref{eq:circular-angular-flux} by the 1PN radial period and consistently
re-expanding gives
\begin{align}
	\Delta E_{\rm circ}
	={}&
	\lambda
	\left[
	-\frac{64\nu\pi}{5\mathcal L^7}
	+
	\varepsilon
	\frac{4\nu\pi(-3289+588\nu)}
	{105\mathcal L^9}
	\right]
	\\&+
	\mathcal O(\lambda\varepsilon^2,\lambda^2),
	\label{eq:circular-energy-loss}
	\\
	\Delta J_{\rm circ}
	={}&
	\lambda
	\left[
	-\frac{64\nu\pi}{5\mathcal L^4}
	+
	\varepsilon
	\frac{4\nu\pi(-1777+756\nu)}
	{105\mathcal L^6}
	\right]
	\\&+
	\mathcal O(\lambda\varepsilon^2,\lambda^2).
	\label{eq:circular-angular-loss}
\end{align}
These expressions coincide with the \(e\rightarrow0\) limit of
Eqs.~\eqref{eq:energy-change-from-magnusian} and
\eqref{eq:angular-change-from-magnusian}.

\subsection{Radial quasi-Keplerian elements}
\label{subsec:radial-elements}

We perform an independent check for \(0<e<1\), without an eccentricity
expansion, by computing the one-cycle changes of the 1PN radial
quasi-Keplerian elements \(a_r\) and \(e_r\) in two ways. First, we transform
the Magnusian kick from \((\mathcal L,\mathcal G)\) to \((a_r,e_r)\). Second,
we multiply the independently known time averages
\(\langle\dot a_r\rangle\) and \(\langle\dot e_r\rangle\) by the 1PN radial
period. In terms of the normal-form Hamiltonian and angular action, the
required elements are \cite{AIHPA_1985__43_1_107_0,Memmesheimer:2004cv}
\begin{align}
	a_r(\mathcal L,\mathcal G)
	&=
	-\frac{1}{2\overline H}
	\left[
	1+
	\varepsilon
	\frac{7-\nu}{2}\overline H
	\right]
	+
	\mathcal O(\varepsilon^2),
	\label{eq:radial-semi-major-axis-actions}
	\\
	e_r^2(\mathcal L,\mathcal G)
	&=
	1+
	2\overline H\mathcal G^2
	\nonumber\\
	&\quad
	+
	\varepsilon\overline H
	\left[
	2(\nu-6)
	+
	5(\nu-3)
	\overline H\mathcal G^2
	\right]
	+
	\mathcal O(\varepsilon^2).
	\label{eq:radial-eccentricity-actions}
\end{align}
Reparametrizing by the action-based eccentricity \(e\) gives
\begin{align}
	a_r
	&=
	\mathcal L^2
	+
	\varepsilon
	\left(
	2-\frac{6}{s}
	\right)
	+
	\mathcal O(\varepsilon^2),
	\label{eq:radial-semi-major-axis-map}
	\\
	e_r
	&=
	e
	+
	\varepsilon
	\frac{
		e\left[6-\nu(1+s)\right]
	}{
		2\mathcal L^2(1+s)
	}
	+
	\mathcal O(\varepsilon^2).
	\label{eq:radial-eccentricity-map}
\end{align}

To display explicitly the Newtonian and relative-1PN contributions, we write
\begin{equation} 	\label{eq:qk-jacobian-expansion}
	\mathsf J^{\rm qK}
	\equiv
	\frac{\partial(a_r,e_r)}
	{\partial(\mathcal L,\mathcal G)}
	=
	\mathsf J_{\rm N}^{\rm qK}
	+
	\varepsilon\mathsf J_{\rm 1PN}^{\rm qK}
	+
	\mathcal O(\varepsilon^2).
\end{equation}
In terms of the PN components of the Jacobian \eqref{eq:qk-jacobian-expansion}, the changes generated by the Magnusian are
\begin{align}
	&\begin{pmatrix}
		\Delta a_r\\
		\Delta e_r
	\end{pmatrix}_{\!\chi}
	=
	\mathsf J_{\rm N}^{\rm qK}
	\begin{pmatrix}
		\chi_{\ell,{\rm N}}^{(1)}\\
		\chi_{g,{\rm N}}^{(1)}
	\end{pmatrix}
	\nonumber\\
	&\quad
	+
	\varepsilon
	\left[
	\mathsf J_{\rm N}^{\rm qK}
	\begin{pmatrix}
		\chi_{\ell,{\rm 1PN}}^{(1)}\\
		\chi_{g,{\rm 1PN}}^{(1)}
	\end{pmatrix}
	+
	\mathsf J_{\rm 1PN}^{\rm qK}
	\begin{pmatrix}
		\chi_{\ell,{\rm N}}^{(1)}\\
		\chi_{g,{\rm N}}^{(1)}
	\end{pmatrix}
	\right]
	\nonumber\\
	&\quad
	+
	\mathcal O(\lambda\varepsilon^2,\lambda^2).
	\label{eq:radial-element-kick}
\end{align}

Independently, the same one-cycle changes can be inferred from the known
orbit-averaged secular rates. Denoting this prediction by the subscript
``av'', one has
\begin{equation}
	\begin{pmatrix}
		\Delta a_r\\
		\Delta e_r
	\end{pmatrix}_{\!\rm av}
	=
	T_r
	\begin{pmatrix}
		\left\langle\dot a_r\right\rangle\\
		\left\langle\dot e_r\right\rangle
	\end{pmatrix}.
\end{equation}
Expanding both the rates and the radial period through relative 1PN order
gives
\begin{align}
	&\begin{pmatrix}
		\Delta a_r\\
		\Delta e_r
	\end{pmatrix}_{\!\rm av}
	=
	T_{r,{\rm N}}
	\begin{pmatrix}
		\left\langle\dot a_r\right\rangle_{\rm N}\\
		\left\langle\dot e_r\right\rangle_{\rm N}
	\end{pmatrix}
	\nonumber\\
	&\quad
	+
	\varepsilon
	\left[
	T_{r,{\rm N}}
	\begin{pmatrix}
		\left\langle\dot a_r\right\rangle_{\rm 1PN}\\
		\left\langle\dot e_r\right\rangle_{\rm 1PN}
	\end{pmatrix}
	+
	T_{r,{\rm 1PN}}
	\begin{pmatrix}
		\left\langle\dot a_r\right\rangle_{\rm N}\\
		\left\langle\dot e_r\right\rangle_{\rm N}
	\end{pmatrix}
	\right]
	\nonumber\\
	&\quad
	+
	\mathcal O(\lambda\varepsilon^2,\lambda^2).
	\label{eq:radial-element-flux-check}
\end{align}
The Newtonian rates are the Peters--Mathews results
\cite{Peters:1963ux,Peters:1964zz}, while their relative-1PN corrections are
taken from
Refs.~\cite{Wagoner:1976am,Blanchet:1989cu,Junker:1992kle,
	Gopakumar:1997bs}. Substituting these rates and the radial-period expansion
of Eq.~\eqref{eq:radial-period-1pn} into
Eq.~\eqref{eq:radial-element-flux-check}, and expressing the Magnusian result
\eqref{eq:radial-element-kick} in the same variables by eliminating
\((\mathcal L,\mathcal G)\) in favor of \((a_r,e_r)\), we find
\begin{equation}
	\begin{pmatrix}
		\Delta a_r\\
		\Delta e_r
	\end{pmatrix}_{\!\chi}
	=
	\begin{pmatrix}
		\Delta a_r\\
		\Delta e_r
	\end{pmatrix}_{\!\rm av}
	+
	\mathcal O(\lambda\varepsilon^2,\lambda^2).
	\label{eq:radial-element-balance-check}
\end{equation}
This equality holds coefficient by coefficient and provides an independent
check of the two Magnusian components that determine the action changes.


\section{Radiation-reaction gauge independence}
\label{sec:gauge}

The calculation in Sec.~\ref{sec:one-cycle-result} used the ADM
representative of the radiation-reaction force. We now restore the complete
Iyer--Will gauge freedom and show that the full four-component one-cycle
covector is unchanged. Through relative 1PN radiation-reaction order, this
family contains the two leading parameters \((\alpha,\beta)\) and six
additional 3.5PN parameters
\((\delta_1,\ldots,\delta_5,\epsilon_5)\)
\cite{Iyer:1993xi,Iyer:1995rn,Nissanke:2004er,Bini:2026suo}.

\subsection{Complete Iyer--Will family}
\label{subsec:complete-iyer-will-family}

We collect the eight gauge parameters into
\begin{equation}
	\bm\gamma
	\equiv
	\left(
	\alpha,\beta,
	\delta_1,\delta_2,\delta_3,
	\delta_4,\delta_5,\epsilon_5
	\right).
	\label{eq:iyer-will-gauge-vector}
\end{equation}
The leading functions \(A_{\rm 2.5PN}\) and \(B_{\rm 2.5PN}\) were given in
Eqs.~\eqref{eq:A25}--\eqref{eq:B25}. In velocity variables, the 3.5PN
functions take the form
\begin{align}
	A_{\rm 3.5PN}
	={}&
	c_1 v^4
	+
	c_2\frac{v^2}{r}
	+
	c_3 v^2\dot r^2
	\nonumber\\
	&\quad
	+
	c_4\frac{\dot r^2}{r}
	+
	c_5\dot r^4
	+
	\frac{c_6}{r^2},
	\label{eq:A35-general}
	\\
	B_{\rm 3.5PN}
	={}&
	d_1 v^4
	+
	d_2\frac{v^2}{r}
	+
	d_3 v^2\dot r^2
	\nonumber\\
	&\quad
	+
	d_4\frac{\dot r^2}{r}
	+
	d_5\dot r^4
	+
	\frac{d_6}{r^2}.
	\label{eq:B35-general}
\end{align}
The twelve coefficients \(c_i\) and \(d_i\) are affine functions of
\(\bm\gamma\), with additional dependence on \(\nu\). Their standard
Iyer--Will expressions are taken from
Refs.~\cite{Iyer:1993xi,Iyer:1995rn,Nissanke:2004er,Bini:2026suo} and
converted to the canonical conventions described below.

The representative used to obtain
Eqs.~\eqref{eq:nlo-magnusian-components} is specified by
\begin{equation}
	\alpha_{\rm ADM}
	=
	\frac{5}{3},
	\qquad
	\beta_{\rm ADM}
	=
	3,
	\label{eq:adm-gauge-25}
\end{equation}
and
\begin{align}
	\delta_{1}^{\rm ADM}
	&=
	\frac{41}{84}
	+
	\frac{677}{42}\nu,
	&
	\delta_{2}^{\rm ADM}
	&=
	-\frac{61}{14}
	-
	\frac{151}{14}\nu,
	\nonumber\\
	\delta_{3}^{\rm ADM}
	&=
	\frac{583}{28}
	-
	\frac{157}{21}\nu,
	&
	\delta_{4}^{\rm ADM}
	&=
	\frac{115}{28}
	-
	\frac{15}{7}\nu,
	\nonumber\\
	\delta_{5}^{\rm ADM}
	&=
	-\frac{961}{42}
	+
	\frac{16}{3}\nu,
	&
\epsilon_5^{\rm ADM}
	&=
	-\frac{51}{14}
	-
	\frac{23}{14}\nu.
	\label{eq:adm-gauge-35}
\end{align}
These ADM values are given in Ref.~\cite{Bini:2026suo}.
Substitution of Eqs.~\eqref{eq:adm-gauge-25} and
\eqref{eq:adm-gauge-35} into the general force reproduces exactly the ADM
force used in Sec.~\ref{sec:one-cycle-result}.

\subsection{Conversion to canonical generalized forces}
\label{subsec:canonical-iyer-will-force}

The Iyer--Will formulas give a relative acceleration as a function of
positions and velocities. The Magnusian construction instead requires the
generalized forces appearing in the canonical ADM momentum equations and,
through Eq.~\eqref{eq:dissipative-covector-lowering}, the associated
doubled-space covector. Converting the acceleration to canonical generalized
forces introduces the conservative
velocity-momentum relation and its Legendre matrix. These corrections first
matter at relative 1PN order.

Let \(p_i(q,v)\) be the conservative canonical momentum obtained from the
1PN Lagrangian. At linear order in the
nonconservative acceleration, its contribution to \(\dot p_i\) is obtained by
differentiating \(p_i\) with respect to the velocity. The corresponding
generalized force is therefore \cite{Bini:2026suo}
\begin{equation}
	\mathcal F_i
	=
	\frac{\partial p_i}{\partial v^j}
	a_{\rm RR}^j.
	\label{eq:canonical-force-from-acceleration}
\end{equation}
We expand the velocity-momentum relation and its Jacobian as
\begin{align}
	v^i(q,p)
	&=
	v_{\rm N}^i(q,p)
	+
	\varepsilon v_{\rm 1PN}^i(q,p)
	+
	\mathcal O(\varepsilon^2),
	\\
	\frac{\partial p_i}{\partial v^j}
	&=
	\delta_{ij}
	+
	\varepsilon\mathsf W^{\rm 1PN}_{ij}
	+
	\mathcal O(\varepsilon^2).
\end{align}
Both quantities are fixed uniquely by the conservative ADM Hamiltonian:
\begin{equation}
	v^i
	=
	\frac{\partial H_{\rm cons}}{\partial p_i},
	\qquad
	\frac{\partial p_i}{\partial v^j}
	=
	\left[
	\frac{\partial^2 H_{\rm cons}}
	{\partial p_i\partial p_j}
	\right]^{-1}.
	\label{eq:velocity-legendre-from-hamiltonian}
\end{equation}
Their 1PN coefficients therefore follow directly from
Eq.~\eqref{eq:adm-hamiltonian-expansion}.

The canonical 2.5PN force is consequently
\begin{equation}
	\mathcal F_i^{\rm 2.5PN}
	=
	a_{{\rm RR},i}^{\rm 2.5PN}
	\big|_{v=v_{\rm N}},
	\label{eq:canonical-force-25}
\end{equation}
whereas the 3.5PN force contains three contributions,
\begin{align}
	&\mathcal F_i^{\rm 3.5PN}
	={}
	a_{{\rm RR},i}^{\rm 3.5PN}
	\big|_{v=v_{\rm N}}
	+
	\left.
	\frac{\partial a_{{\rm RR},i}^{\rm 2.5PN}}
	{\partial v^j}
	\right|_{v=v_{\rm N}}
	v_{\rm 1PN}^j
	\nonumber\\
	&\quad
	+
	\mathsf W^{\rm 1PN}_{ij}
	a_{{\rm RR},j}^{\rm 2.5PN}
	\big|_{v=v_{\rm N}}.
	\label{eq:canonical-force-35}
\end{align}
The second term accounts for the 1PN inversion of the
velocity-momentum relation, while the third comes from the 1PN Legendre
matrix in Eq.~\eqref{eq:canonical-force-from-acceleration}. Both are required
for a consistent canonical 3.5PN force.

The resulting canonical generalized force has the polar components obtained
from Eq.~\eqref{eq:generalized-polar-force}. Its symplectic lowering is the
doubled-space covector in Eq.~\eqref{eq:polar-force-one-form}. Applying
Eqs.~\eqref{eq:canonical-force-25}--\eqref{eq:canonical-force-35} to the
general Iyer--Will acceleration of
Refs.~\cite{Iyer:1993xi,Iyer:1995rn,Nissanke:2004er,Bini:2026suo}
reproduces the general 2.5PN canonical force and, after imposing
Eqs.~\eqref{eq:adm-gauge-25}--\eqref{eq:adm-gauge-35}, the ADM 3.5PN force
used above.

\subsection{Direct computation of the gauge-response matrix}
\label{subsec:gauge-response}

To test the complete Iyer--Will family, we repeat the construction of the
1PN pullback integrand without imposing the ADM values
\(\bm\gamma_{\rm ADM}\). More precisely, the general canonical forces obtained
from Eqs.~\eqref{eq:canonical-force-25} and
\eqref{eq:canonical-force-35} are inserted into all four terms of
Eq.~\eqref{eq:nlo-pullback-integrand}. This produces the general-gauge
integrand~\footnote{The dependence on \(\alpha\) and \(\beta\) is included both
	through the leading 2.5PN force and through their appearance in the 3.5PN
	coefficients.}
\begin{equation}
	\mathcal I_{{\rm 1PN},A}^{(1)}
	(u;\bm\gamma),
	\qquad
	A\in\{\ell,g,\mathcal L,\mathcal G\}.
	\label{eq:general-gauge-nlo-integrand}
\end{equation}

Each force coefficient is a parameter-independent term plus a term linear in
the eight entries of $\bm\gamma$; no products of gauge parameters occur.
Consequently, the complete 1PN integrand admits the exact decomposition
\begin{align}
	&	\mathcal I_{{\rm 1PN},A}^{(1)}(u;\bm\gamma)
	=
	\mathcal I_{{\rm 1PN},A}^{(1)}
	(u;\bm\gamma_{\rm ADM})
	\nonumber\\&\qquad+
	\sum_{I=1}^{8}
	\left(
	\gamma_I-\gamma_I^{\rm ADM}
	\right)
	\mathcal R_{IA}(u),
	\label{eq:gauge-integrand-decomposition}
\end{align}
where
\begin{equation}
	\mathcal R_{IA}(u)
	\equiv
	\frac{\partial
		\mathcal I_{{\rm 1PN},A}^{(1)}(u;\bm\gamma)}
	{\partial\gamma_I}.
	\label{eq:gauge-response-integrands}
\end{equation}
The calculation therefore produces 32 gauge-response integrands:
one for each of the eight parameters and each of the four Magnusian
components. These functions need not vanish pointwise.

The corresponding response of the one-cycle covector is the \(8\times4\)
matrix
\begin{equation}
	\mathsf G_{IA}
	\equiv
	\frac{\partial\chi_{{\rm 1PN},A}^{(1)}}
	{\partial\gamma_I}
	=
	\int_0^{2\pi}\!\diff u\,
	\mathcal R_{IA}(u).
	\label{eq:gauge-response-matrix-definition}
\end{equation}
We evaluate every entry of this matrix exactly. First, the two halves of the
radial cycle are combined:
\begin{equation}
	\mathcal R^{\rm pair}_{IA}(u)
	\equiv
	\mathcal R_{IA}(u)
	+
	\mathcal R_{IA}(2\pi-u),
	\qquad
	0\leq u\leq\pi.
	\label{eq:paired-gauge-response}
\end{equation}
It follows that
\begin{equation}
	\mathsf G_{IA}
	=
	\int_0^\pi\!\diff u\,
	\mathcal R^{\rm pair}_{IA}(u).
	\label{eq:paired-gauge-response-integral}
\end{equation}

We then introduce the half-angle variable \(x=\tan(u/2)\) and define
\begin{equation}
	\widehat{\mathcal R}_{IA}(x)
	\equiv
	\left.
	\mathcal R^{\rm pair}_{IA}(u)
	\frac{\diff u}{\diff x}
	\right|_{u=2\arctan x}.
	\label{eq:rational-gauge-response}
\end{equation}
After exact algebraic reduction, every
\(\widehat{\mathcal R}_{IA}\) is a rational function of \(x\), with
coefficients depending on
\((e,s,\mathcal L,\nu)\). Hence,
\begin{equation}
	\mathsf G_{IA}
	=
	\int_0^\infty\!\diff x\,
	\widehat{\mathcal R}_{IA}(x).
	\label{eq:rational-gauge-response-integral}
\end{equation}
Evaluating these integrals with the same beta-function reduction used in
Sec.~\ref{subsec:exact-radial-integration} gives
\begin{equation}
	\mathsf G_{IA}
	=
	0,
	\qquad
	I=1,\ldots,8,
	\qquad
	A\in\{\ell,g,\mathcal L,\mathcal G\}.
	\label{eq:complete-gauge-response-zero}
\end{equation}

Substituting this result into the integrated version of
Eq.~\eqref{eq:gauge-integrand-decomposition} gives
\begin{align}
	\chi_{{\rm 1PN},A}^{(1)}(\bm\gamma)
	&=
	\chi_{{\rm 1PN},A}^{(1)}(\bm\gamma_{\rm ADM})
	+
	\sum_{I=1}^{8}
	\left(
	\gamma_I-\gamma_I^{\rm ADM}
	\right)
	\mathsf G_{IA}
	\nonumber\\
	&=
	\chi_{{\rm 1PN},A}^{(1)}(\bm\gamma_{\rm ADM}).
		\label{eq:general-gauge-reconstruction}
\end{align}
This establishes independence from all eight Iyer--Will gauge parameters for
all four components, not only for the two physical action losses.

The analogous leading-order calculation starts from the general
\((\alpha,\beta)\)-dependent 2.5PN pullback. Its \(2\times4\) response matrix
also vanishes:
\begin{equation}
	\frac{\partial\chi_{{\rm N},A}^{(1)}}{\partial\alpha}
	=
	\frac{\partial\chi_{{\rm N},A}^{(1)}}{\partial\beta}
	=
	0,
	\qquad
	A\in\{\ell,g,\mathcal L,\mathcal G\}.
		\label{eq:leading-gauge-independence}
\end{equation}
Consequently, the result in
Sec.~\ref{subsec:closed-form-magnusian} represents the complete
eight-parameter Iyer--Will family through relative 1PN order.

The vanishing response can be understood as follows: after pairing the two halves of the orbit and applying the
half-angle substitution, each of the 32 gauge-response integrands is an exact
derivative whose primitive vanishes at both endpoints. The proof is given in
Appendix~\ref{app:periodic-coboundary}.


\section{Relation to Blanco's leading-order result}
\label{sec:blanco-reduction}

We now compare our result with the leading-order construction of
Ref.~\cite{Blanco:2026prj}. The comparison fixes the precise overlap of the
two calculations and distinguishes the strict first-order map from the partially exponentiated map generated by the first Magnusian.

\subsection{Leading-order reduction}
\label{subsec:blanco-leading-reduction}

Blanco uses the harmonic-coordinate representative of the leading
radiation-reaction force. In the Iyer--Will parametrization employed here, it
is obtained by setting
\begin{equation}
	\alpha=-1,
	\qquad
	\beta=0.
	\label{eq:blanco-harmonic-gauge}
\end{equation}
This choice reproduces Eq.~(4.2) of Ref.~\cite{Blanco:2026prj}. Gauge
independence of the one-cycle integral then guarantees that it may be compared
directly with our ADM calculation.

Setting $\varepsilon=0$ and expressing the result in terms of the
action-based eccentricity, our leading coefficient becomes~\footnote{For consistency with Blanco's notation, we set $\lambda=1$ here.}
\begin{subequations}
	\begin{align}
		\chi_{\ell,\rm N}^{(1)}
		&=
		-\frac{2\pi\nu\mathcal L^3}{15\mathcal G^7}
		\left(
		96+292e^2+37e^4
		\right),
		\label{eq:blanco-reduction-chi-l}
		\\
		\chi_{g,\rm N}^{(1)}
		&=
		-\frac{8\pi\nu}{5\mathcal G^4}
		\left(
		8+7e^2
		\right),
		\label{eq:blanco-reduction-chi-g}
		\\
		\chi_{\mathcal L,\rm N}^{(1)}
		&=
		\frac{2\pi^2\nu\mathcal L^2}{5\mathcal G^7}
		\left(
		96+292e^2+37e^4
		\right),
		\label{eq:blanco-reduction-chi-L}
		\\
		\chi_{\mathcal G,\rm N}^{(1)}
		&=0.
		\label{eq:blanco-reduction-chi-G}
	\end{align}
\end{subequations}
All the components agree with Eqs.~(4.21a)--(4.21d) of Ref.~\cite{Blanco:2026prj}.



\subsection{Strict map and partial exponentiation}
\label{subsec:strict-versus-partial-exponential}

With $\mathcal D_1$ already defined in
Eq.~\eqref{eq:magnus-derivation-definition}, the strict first-order map is
Eq.~\eqref{eq:first-order-observable-map}, namely the
$\mathcal O(\lambda)$ truncation of the common Magnus exponential
\eqref{eq:formal-magnus-exponential}. We denote this prescription by
``str''. Ref.~\cite{Blanco:2026prj} instead uses partial exponentiation,
denoted by ``pe'', retaining the first three nested
actions of the same $\mathcal D_1$,
\begin{align}
	\mathcal O_f^{\rm pe}
	={}&
	\mathcal O
	+
	\lambda\mathcal D_1\mathcal O
	+
	\frac{\lambda^2}{2}
	\mathcal D_1^2\mathcal O
	\nonumber\\
	&+
	\frac{\lambda^3}{6}
	\mathcal D_1^3\mathcal O
	+
	\mathcal O(\lambda^4).
	\label{eq:blanco-partial-exponential}
\end{align}
The two maps therefore agree through the order derived analytically here.
To display precisely which terms are selected, the complete map through
cubic order is
\begin{align}
	\mathcal O_f
	={}&
	\mathcal O
	+
	\lambda\mathcal D_1\mathcal O
	+
	\lambda^2
	\left(
	\mathcal D_2+\frac12\mathcal D_1^2
	\right)\mathcal O
	\nonumber\\
	&+
	\lambda^3
	\left[
	\mathcal D_3
	+\frac12
	\left(
	\mathcal D_1\mathcal D_2+
	\mathcal D_2\mathcal D_1
	\right)
	\right.
	\nonumber\\
	&\qquad\left.
	+\frac16\mathcal D_1^3
	\right]\mathcal O
	+
	\mathcal O(\lambda^4).
	\label{eq:complete-map-through-cubic}
\end{align}
Thus the partially exponentiated map retains the $\mathcal D_1^2/2$ and
$\mathcal D_1^3/6$ pieces, but omits $\mathcal D_2$ at quadratic order and
$\mathcal D_3$ together with the mixed
$\mathcal D_1\mathcal D_2$ terms at cubic order. It is therefore a
well-defined resummation of the known first Magnusian, which should not be confused with the complete second- or third-order Magnus map.


\section{Numerical comparison of the map prescriptions}
\label{sec:numerical-comparison}

Having established the first Magnusian analytically through relative 1PN
order, we use the numerical calculations below to compare two finite-map
prescriptions against a specified target: the non-re-expanded solution of the
PN-truncated ADM equations. We first examine accumulated tracking of both the
stroboscopic state and the periastron arrival time for one representative
orbit and then test the one-cycle state-space comparison across orbital
parameter space. The formal bookkeeping parameters are set to
\(\lambda=\varepsilon=1\) throughout this section.
A short formal scaling diagnostic in which \(\lambda\) is varied independently
is reported in Appendix~\ref{app:formal-lambda-diagnostic}.

\subsection{Map prescriptions, direct evolution, and residuals}
\label{subsec:definition-numerical-comparison}

Let \(X\in\{{\rm str},{\rm pe}\}\) label the two prescriptions under scrutiny.
Their raw one-cycle finite-map update has the common form
\begin{subequations}
	\begin{align}
		\bm Y_{n+1}^{X}
		&=
		\left.
		\mathcal M_X\bm Y
		\right|_{\bm Y=\bm Y_n^{X}},
		\label{eq:common-stroboscopic-map}
		\\
		t_{n+1}^{X}
		&=
		t_n^{X}
		+
		T_r(\bm J_n^{X})
		+
		\frac{\Delta\ell_{{\rm RR},n}^{X}}
		{\omega_\ell(\bm J_n^{X})},
		\label{eq:stroboscopic-time-map}
		\\
		\Delta\ell_{{\rm RR},n}^{X}
		&\equiv
		\left[
		\mathcal M_X\bm Y_n^{X}
		\right]_{\ell}
		-
		\ell_n^{X}
		-
		2\pi.
		\label{eq:stroboscopic-mean-anomaly-offset}
	\end{align}
\end{subequations}
The quantity $\bm Y_{n+1}^{X}$ in
Eq.~\eqref{eq:common-stroboscopic-map} denotes the finite Magnus update before
projection onto the perturbed return section. At the accuracy controlled by
the first Magnusian, this distinction does not affect the action components,
whereas the angle sector must be converted to the corresponding
return-section quantities according to
Sec.~\ref{subsec:return-section}.

Eq.~\eqref{eq:common-stroboscopic-map} applies to both the
\({\rm str}\) and \({\rm pe}\) prescriptions
once the corresponding operator $\mathcal M_X$ is specified. For
${\rm str}$, the update within each radial cycle is truncated at first order
in $\lambda$; repeated iteration changes the orbit on which the subsequent
kick is evaluated but does not alter this within-cycle truncation. For
${\rm pe}$, the second and third nested actions generated by the same
first-order covector are retained, consistently re-expanded through first
order in $\varepsilon$, before each one-cycle update.

The time update in Eq.~\eqref{eq:stroboscopic-time-map} implements the
corresponding projection onto the next periastron section. Its first term is
the conservative radial period evaluated on the current actions, while the
second is the return-time correction associated with the radial-phase
displacement generated by the finite map. For the strict prescription this
term is precisely the first-order correction derived in
Eq.~\eqref{eq:return-time-correction}. For the partially exponentiated map,
the same construction is applied to the resummed radial-phase update and
therefore retains selected higher-order contributions in $\lambda$.

For comparison, we directly integrate the local ADM equations
\begin{subequations}
	\label{eq:direct-adm-coordinate-equations}
	\begin{align}
		\dot r
		&=
		\frac{\partial H_{\rm cons}}{\partial p_r},
		\qquad
		\dot\phi=
		\frac{\partial H_{\rm cons}}{\partial p_\phi},
		\\
		\dot p_r
		&=
		-\frac{\partial H_{\rm cons}}{\partial r}
		+
		\mathcal F_r^{\rm 2.5PN}
		+
		\mathcal F_r^{\rm 3.5PN}
		,
		\\
		\dot p_\phi
		&=
		\mathcal F_\phi^{\rm 2.5PN}
		+
		\mathcal F_\phi^{\rm 3.5PN}
		.
	\end{align}
\end{subequations}
Successive periastra are located as outward crossings of $p_r=0$, and the
continuous trajectory is sampled only at those events. The direct integration
used for the representative checkpoint was evaluated with Wolfram
Mathematica \cite{Mathematica_14_3} using \texttt{NDSolveValue} with its automatic integration
and event-location methods, 40-digit working precision, accuracy and precision
goals of 24 and 20 digits, respectively, no explicit maximum step size, and
\texttt{MaxSteps -> Infinity}. The event condition is implemented with
\texttt{WhenEvent} at $p_r=0$ and positive crossing direction.

The parameter grid below is obtained using Python~3.12 and
SciPy~1.17 \cite{2020SciPy-NMeth}. The direct equations are integrated
with \texttt{solve\_ivp} and the DOP853 method
\cite{HairerNorsettWanner_1993}, relative and absolute
tolerances $2\times10^{-13}$ and $2\times10^{-15}$, and maximum step
$T_r(\bm J_0)/120$; periastra are positive-direction roots of $p_r$ located by
the solver's event finder. At the grid point
$(p_0,e_0)=(50,0.3)$, the four one-cycle diagnostics agree with the
Mathematica checkpoint to within $3.6\times10^{-12}$. Varying either the
relative tolerance by a factor of 10 or the maximum step by a factor of 2
changes every grid value of the residual ratio defined below by less than
$1.4\times10^{-6}$.

Here and below, the label ``${\rm dir}$'' denotes quantities extracted from the
direct integration of Eqs.~\eqref{eq:direct-adm-coordinate-equations}. At each periastron of the direct evolution, we set
\begin{equation}
	\mathcal G_{\rm dir}=p_\phi,
	\qquad
	E_{\rm dir}
	=
	H_{\rm cons}(r,\phi,p_r,p_\phi),
	\label{eq:direct-invariants}
\end{equation}
and determine $\mathcal L_{\rm dir}$ by solving
\begin{equation}
	\overline H
	\left(
	\mathcal L_{\rm dir},
	\mathcal G_{\rm dir}
	\right)
	=
	E_{\rm dir},
	\label{eq:direct-L-reconstruction}
\end{equation}
through 1PN order. The corresponding action-based eccentricity is then
$
e_{\rm dir}
\equiv
e(\mathcal L_{\rm dir},
\mathcal G_{\rm dir})
$.
For any
$Q\in\{\mathcal L,\mathcal G,e,E\}$, we define the periastron residual by
\begin{equation}
	\delta Q_n
	\equiv
	Q_n^{\rm map}-Q_n^{\rm dir}.
\end{equation}
The two maps and the direct evolution are therefore initialized and compared
using the same normal-form invariants.

\subsection{Finite-strength stroboscopic comparison}
\label{subsec:finite-strength-comparison}

We consider the representative equal-mass configuration~\footnote{Here $p_0$ is the Newtonian semilatus rectum of the initial orbit.}
\begin{equation}
	\nu=\frac14,
	\qquad
	e_0=0.3,
	\qquad
	p_0\equiv\mathcal G_0^2=50,
	\label{eq:numerical-checkpoint}
\end{equation}
with
\begin{equation}
	\mathcal L_0
	=
	\sqrt{\frac{p_0}{1-e_0^2}},
	\qquad
	\ell_0=g_0=0.
\end{equation}
The direct trajectory is initialized at periastron by setting
\begin{equation}
	\phi_0=0,
	\qquad
	p_{r0}=0,
	\qquad
	p_{\phi0}=\mathcal G_0=5\sqrt{2}.
\end{equation}
The remaining initial datum, the periastron radius $r_0$, is fixed by matching
the conservative energy of the direct ADM trajectory to that of the
normal-form orbit. We therefore solve
\begin{equation}
	H_{\rm cons}(r_0,0,0,\mathcal G_0)
	=
	\overline H(\mathcal L_0,\mathcal G_0),
	\label{eq:initial-invariant-matching}
\end{equation}
selecting the branch that reduces to the Newtonian periastron
$r_{p,{\rm N}}=\mathcal L_0^2(1-e_0)$ as
$\varepsilon\to0$. This gives
\begin{equation}
	\bm z_0
	=
	\left(
	34.2193,\,
	0,\,
	0,\,
	5\sqrt{2}
	\right),
	\label{eq:direct-initial-state}
\end{equation}
where the displayed value of $r_0$ is rounded.

The comparison in this subsection pairs the $n$th map iterate with the $n$th
periastron of the direct evolution. It therefore tests the stroboscopic state
at equivalent radial sections; the accumulated arrival times of those
sections are examined separately below.

\begin{figure*}[t]
	\centering
	\includegraphics[width=\textwidth]
	{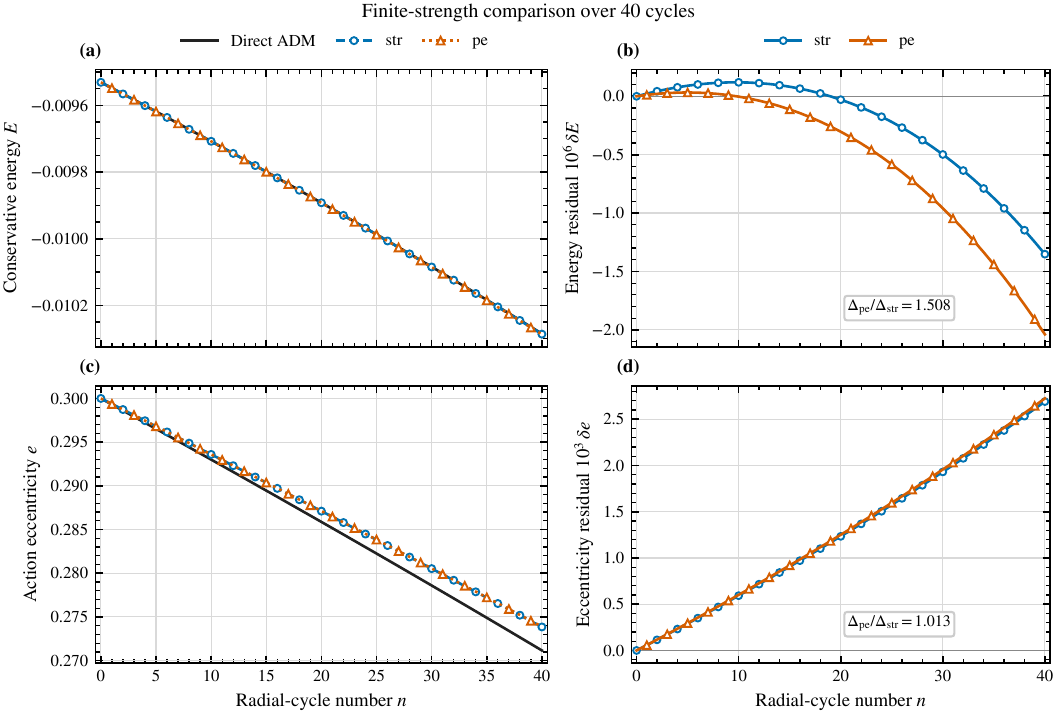}
	\caption{
		Finite-strength comparison for the configuration
		\eqref{eq:numerical-checkpoint}. Panels (a) and (c) show the conservative
		energy and action-based eccentricity at successive periastra, while panels
		(b) and (d) show the corresponding residuals
		\(\delta Q_n=Q_n^{\rm map}-Q_n^{\rm dir}\). The direct ADM sequence is
		black, the \({\rm str}\) map is blue with circles, and the \({\rm pe}\)
		map is orange with triangles. The trajectory curves remain close, while
		the residual panels resolve the accumulated offsets at equal periastron
		index.
	}
	\label{fig:finite-strength-trajectories}
\end{figure*}

Both maps are iterated for 40 radial cycles. For the strict truncation, the
maximum absolute residuals over $0\leq n\leq40$ are
\begin{subequations}
	\label{eq:finite-strength-residuals}
	\begin{align}
		\max_n |\delta\mathcal L_n|_{\rm str}
		&=1.4411\times10^{-4},
		\\
		\max_n |\delta\mathcal G_n|_{\rm str}
		&=5.5740\times10^{-3},
		\\
		\max_n |\delta e_n|_{\rm str}
		&=2.6911\times10^{-3},
		\\
		\max_n |\delta E_n|_{\rm str}
		&=1.3521\times10^{-6}.
	\end{align}
\end{subequations}
For the partially exponentiated map, the corresponding values are
\begin{subequations}
	\label{eq:finite-strength-partial-residuals}
	\begin{align}
		\max_n |\delta\mathcal L_n|_{\rm pe}
		&=3.5749\times10^{-4},
		\\
		\max_n |\delta\mathcal G_n|_{\rm pe}
		&=5.8584\times10^{-3},
		\\
		\max_n |\delta e_n|_{\rm pe}
		&=2.7252\times10^{-3},
		\\
		\max_n |\delta E_n|_{\rm pe}
		&=2.0387\times10^{-6}.
	\end{align}
\end{subequations}
In the ordering $(\mathcal L,\mathcal G,e,E)$, the ratios between the two sets
of maxima are
\begin{equation}
	\frac{\max|\delta Q|_{\rm pe}}
	{\max|\delta Q|_{\rm str}}
	=
	(2.481,\,1.051,\,1.013,\,1.508).
	\label{eq:finite-strength-partial-strict-ratios}
\end{equation}
For this orbit and evolution interval, the strict prescription therefore gives
the smaller maximum stroboscopic residual in all four diagnostics. The separation is most
pronounced for $\mathcal L$ and $E$, while the two prescriptions remain much
closer for $\mathcal G$ and $e$. Fig.~\ref{fig:finite-strength-trajectories}
shows that the energy sequences are visually indistinguishable on the
trajectory scale and that the eccentricity curves remain close, whereas the
residual panels clearly resolve their relative ordering.

The smooth accumulated residuals have a direct dynamical interpretation. The
${\rm str}$ map evaluates the complete first-order kick along a conservative
orbit and applies it at the end of each radial cycle. In the direct evolution,
the actions instead change continuously within the cycle and feed back into
the local radiation-reaction force. This within-cycle feedback is absent from
the strict map and begins at $\mathcal O(\lambda^2)$. Its coherent
cycle-to-cycle accumulation provides a natural explanation for the
approximately monotonic residuals in panels (b) and (d) of
Fig.~\ref{fig:finite-strength-trajectories}.

The ${\rm pe}$ map supplements the first-order kick with the selected nested
contributions $\mathcal D_1^2/2$ and $\mathcal D_1^3/6$. For the present
trajectory, these terms produce the larger state-variable offsets quantified
in Eq.~\eqref{eq:finite-strength-partial-strict-ratios}. The comparison thus
provides an indication of how this particular resummation modifies the
accumulated stroboscopic state relative to the strict first-order prescription.

The direct reference itself is obtained by solving the PN-truncated equations
without re-expanding them in $\varepsilon$. Their numerical integration
therefore generates products proportional to
$\varepsilon^2,\varepsilon^3,\ldots$, defining a specific higher-PN completion
of the retained Newtonian and relative-1PN dynamics. Partial exponentiation
similarly defines a specific higher-$\lambda$ completion through nested actions
of $\mathcal D_1$.

The different visual sensitivities of the energy and eccentricity diagnostics
follow directly from their dependence on the actions. Linearizing about either
trajectory gives
\begin{subequations}
	\label{eq:diagnostic-residual-linearization}
	\begin{align}
		\delta e_n
		&=
		\frac{s_n^2}{e_n\mathcal L_n}\,
		\delta\mathcal L_n
		-
		\frac{s_n}{e_n\mathcal L_n}\,
		\delta\mathcal G_n
		+
		\mathcal O\!\left(\|\delta\bm J_n\|^2\right),
		\\
		\delta E_n
		&=
		\omega_{\ell,n}\delta\mathcal L_n
		+
		\omega_{g,n}\delta\mathcal G_n
		+
		\mathcal O\!\left(\|\delta\bm J_n\|^2\right).
	\end{align}
\end{subequations}
For the initial configuration in Eq.~\eqref{eq:numerical-checkpoint}, the
eccentricity derivatives are $\partial e/\partial\mathcal L=0.409$ and
$\partial e/\partial\mathcal G=-0.429$, whereas
\begin{equation}
	(\omega_{\ell,0},\omega_{g,0})
	=
	(2.55\times10^{-3},\,1.47\times10^{-4}).
\end{equation}
Using the maximal action residuals in
Eq.~\eqref{eq:finite-strength-residuals} as characteristic scales gives
$|\delta e|\sim2.4\times10^{-3}$ and
$|\delta E|\sim1.2\times10^{-6}$, close to the measured maxima. The
eccentricity offset is therefore driven mainly by the angular-momentum
residual, while the same action mismatch is strongly suppressed when
propagated to the energy. In fractional terms,
\begin{equation}
	\frac{\max|\delta e|}{e_0}
	\simeq9.0\times10^{-3},
	\qquad
	\frac{\max|\delta E|}{|E_0|}
	\simeq1.4\times10^{-4}.
\end{equation}

\subsection{Periastron return time and apsidal validation}
\label{subsec:physical-time-dephasing}

The comparison in Sec.~\ref{subsec:finite-strength-comparison} is performed
at equal periastron index and concerns the action-based state variables.
It does not test the time at which each periastron is reached. We first
examine this timing information and then perform a separate check of the
apsidal part of the return-section projection derived in
Sec.~\ref{subsec:return-section}.

For the timing comparison, we consider the strict map together with two
partially exponentiated prescriptions. We denote the three-bracket map of
Eq.~\eqref{eq:blanco-partial-exponential} by ${\rm pe}_3$ and introduce the
auxiliary two-bracket map
\begin{equation}
	\mathcal O_f^{{\rm pe}_2}
	=
	\mathcal O
	+
	\lambda\mathcal D_1\mathcal O
	+
	\frac{\lambda^2}{2}\mathcal D_1^2\mathcal O
	+
	\mathcal O(\lambda^3).
	\label{eq:two-bracket-partial-map}
\end{equation}
For $X\in\{{\rm str},{\rm pe}_2,{\rm pe}_3\}$, we define the periastron-time
residual
\begin{equation}
	\delta t_n^X
	\equiv
	t_n^X-t_n^{\rm dir},
	\label{eq:periastron-time-residual}
\end{equation}
where $t_n^X$ is the return time predicted by the corresponding finite map
through Eq.~\eqref{eq:stroboscopic-time-map}, while $t_n^{\rm dir}$ is the
time of the $n$th periastron obtained from the event-located direct ADM
evolution. For ${\rm str}$, the time update implements the first-order
return-section correction derived in
Eq.~\eqref{eq:return-time-correction}. For ${\rm pe}_2$ and ${\rm pe}_3$,
the same construction is applied to their resummed radial-phase updates;
these prescriptions therefore retain selected higher powers of $\lambda$
without representing complete higher-order Magnus return maps.
To isolate the relative-1PN contribution from the non-re-expanded numerical
evolution, we use the PN projection described in
Appendix~\ref{app:formal-lambda-diagnostic}.

For all the cases considered here, the largest absolute timing residual over
the first \(40\) cycles occurs at \(n=40\). The results are summarized in
Table~\ref{tab:periastron-timing-residuals}. In addition to the representative
configuration \eqref{eq:numerical-checkpoint}, we include the Newtonian orbit
used in Ref.~\cite{Blanco:2026prj},
$h_0\equiv\mathcal G_0=20$ and $e_0=0.3$, corresponding to $p_0=400$ in our
notation.

\begin{table}[t]
	\caption{Absolute periastron-time residual after $40$ cycles,
		$|\delta t_{40}|$, in units of $M$. Here N denotes Newtonian conservative
		dynamics, while N+1PN denotes the result through relative 1PN order.}
	\label{tab:periastron-timing-residuals}
	\scriptsize
	\setlength{\tabcolsep}{2pt}
	\begin{ruledtabular}
		\begin{tabular}{@{}lccc@{}}
			Case & ${\rm str}$ & ${\rm pe}_2$ & ${\rm pe}_3$ \\
			\hline
			N, $p_0=50$
			& $2.9127$ & $0.2430$ & $0.2510$ \\
			N, $h_0=20$
			& $1.6994\times10^{-3}$
			& $1.6747\times10^{-4}$
			& $1.6750\times10^{-4}$ \\
			N+1PN, $p_0=50$
			& $3.7182$ & $0.2957$ & $0.3069$ \\
		\end{tabular}
	\end{ruledtabular}
\end{table}

For the Newtonian $p_0=50$ orbit, ${\rm pe}_2$ and ${\rm pe}_3$ reduce the
accumulated timing residual relative to ${\rm str}$ by factors $11.99$ and
$11.60$, respectively. For the weaker-field $h_0=20$ orbit, both reduction
factors are approximately $10.15$. The same ordering is obtained through
relative 1PN order for $p_0=50$, with reduction factors $12.57$ and $12.12$.

The Newtonian and relative-1PN rows for $p_0=50$ correspond to different
PN-truncated target dynamics and should therefore not be interpreted as a
PN-convergence sequence. Numerically, the inclusion of the relative-1PN
sector increases the accumulated timing residual by approximately $28\%$
for ${\rm str}$ and by approximately $22\%$ for ${\rm pe}_2$ and
${\rm pe}_3$, without changing the ordering of the three prescriptions.

The ordering after a single cycle is instead reversed. For the representative
$p_0=50$ orbit at Newtonian order, we find
\begin{align}
	|\delta t_1^{\rm str}|
	&=
	3.1828\times10^{-3},
	\\
	|\delta t_1^{{\rm pe}_2}|
	&=
	6.0625\times10^{-3},
	\\
	|\delta t_1^{{\rm pe}_3}|
	&=
	6.0689\times10^{-3}.
	\label{eq:first-cycle-timing-residuals}
\end{align}
Thus, the strict map gives the smaller first-cycle timing residual, by a
factor of about $1.9$ relative to either partially exponentiated
prescription. After repeated iteration, however, this ordering reverses:
after $40$ cycles the partially exponentiated maps reduce the accumulated
timing residual by approximately one order of magnitude. The accuracy of a
single return therefore does not by itself predict the timing error
accumulated over many cycles.

The two partially exponentiated prescriptions remain very close throughout
this comparison, with ${\rm pe}_2$ marginally more accurate than
${\rm pe}_3$. The improvement obtained by retaining selected nested actions
is therefore not monotonic in their number and should not be interpreted as
an increase in formal perturbative accuracy.

We finally test the apsidal part of the return-section construction. We
restrict this comparison to the strict map because the apsidal correction
derived in Sec.~\ref{subsec:return-section} is controlled through first order
in $\lambda$. Applying an analogous construction to ${\rm pe}_2$ or
${\rm pe}_3$ would retain selected higher-order terms generated by
$\mathcal D_1$ while omitting the independent higher Magnus generators, and
would therefore not constitute a controlled higher-order apsidal prediction.

The comparison with the direct ADM azimuth is particularly simple at
periastron. Setting the radial anomaly to its periastron value in the
explicit 1PN ADM--Delaunay transformation gives
\begin{equation}
	\phi_{\rm peri}=g
	\qquad
	(\mathrm{mod}\ 2\pi).
	\label{eq:periastron-phi-g-relation}
\end{equation}
Consequently, if $g_n^{\rm str}$ denotes the apsidal angle produced by the
strict return map after $n$ radial cycles, its continuously unwrapped
periastron azimuth is
\begin{equation}
	\phi_{{\rm peri},n}^{\rm str}
	=
	2\pi n+g_n^{\rm str}.
	\label{eq:strict-periastron-azimuth}
\end{equation}
We compare this quantity directly with the continuously evolved ADM azimuth
$\phi_{{\rm peri},n}^{\rm dir}$ evaluated at the $n$th numerically located
periastron, and define
\begin{equation}
	\delta\phi_{{\rm peri},n}^{\rm str}
	\equiv
	\phi_{{\rm peri},n}^{\rm str}
	-
	\phi_{{\rm peri},n}^{\rm dir}.
	\label{eq:periastron-azimuth-residual}
\end{equation}

For the representative $p_0=50$ evolution over $40$ cycles, we obtain
\begin{align}
	\max_{0\leq n\leq40}
	|\delta\phi_{{\rm peri},n}^{\rm str}|_{\rm N}
	&=
	1.6773\times10^{-4},
	\\
	\max_{0\leq n\leq40}
	|\delta\phi_{{\rm peri},n}^{\rm str}|_{\leq{\rm 1PN}}
	&=
	7.0498\times10^{-4}.
	\label{eq:periastron-azimuth-validation}
\end{align}
At Newtonian order the dissipative apsidal correction of
Eq.~\eqref{eq:return-apsidal-shift} vanishes, whereas through relative 1PN
order the strict return map includes the
$\mathcal O(\lambda\varepsilon)$ correction derived in
Eq.~\eqref{eq:return-apsidal-shift-explicit}. The comparison above therefore
provides a direct numerical test of the apsidal sector of the first-order
periastron return map.

\subsection{One-cycle comparison over orbital parameter space}
\label{subsec:orbital-parameter-robustness}

The accumulated stroboscopic and timing comparisons above follow one orbit for
many cycles. We now perform a complementary one-cycle state-space test over
orbital parameter space. At
\(\nu=1/4\), we use
\begin{equation}
	p_0\in\{50,70,100,140\},
	\quad
	e_0\in\{0.1,0.3,0.5,0.7\}.
	\label{eq:orbital-parameter-grid}
\end{equation}
Every direct trajectory is rematched to the same initial normal-form
invariants as the two maps and evolved from one periastron to the next. As a simple indication of the strength of the relative-1PN dissipative
correction over the sampled region, we evaluate at the initial periastron
\begin{equation}
	\mathcal R_\phi
	\equiv
	\left|
	\frac{\mathcal F_\phi^{\rm 3.5PN}}
	{\mathcal F_\phi^{\rm 2.5PN}}
	\right|_{\rm peri}.
\end{equation}
Over the grid \eqref{eq:orbital-parameter-grid}, this ratio ranges from
$0.112$ to $0.606$. Thus, although the relative-1PN correction becomes
sizeable for the most relativistic configurations considered, it remains
smaller than the leading 2.5PN contribution throughout the sampled region.

For \(X\in\{{\rm str},{\rm pe}\}\), define the one-cycle error normalized by
the direct one-cycle change,
\begin{equation}
	\mathcal E_Q^X(p_0,e_0)
	\equiv
	\frac{\left|Q_1^X-Q_1^{\rm dir}\right|}
	{\left|Q_1^{\rm dir}-Q_0\right|},
	\qquad
	Q\in\{\mathcal L,\mathcal G,e,E\}.
	\label{eq:normalized-one-cycle-error}
\end{equation}
The quantity \(\mathcal E_Q^X\) measures the fraction of the direct secular change
missed by one map step. To compare the prescriptions without introducing a
scale for each observable, we use
\begin{equation}
	\rho_Q(p_0,e_0)
	\equiv
	\frac{\mathcal E_Q^{\rm pe}}{\mathcal E_Q^{\rm str}}
	=
	\frac{\left|\delta Q_1\right|_{\rm pe}}
	{\left|\delta Q_1\right|_{\rm str}}.
	\label{eq:one-cycle-error-ratio}
\end{equation}
Values \(\rho_Q<1\) favor \({\rm pe}\), whereas \(\rho_Q>1\) favor
\({\rm str}\).

\begin{figure*}[t]
	\centering
	\includegraphics[width=\textwidth]
	{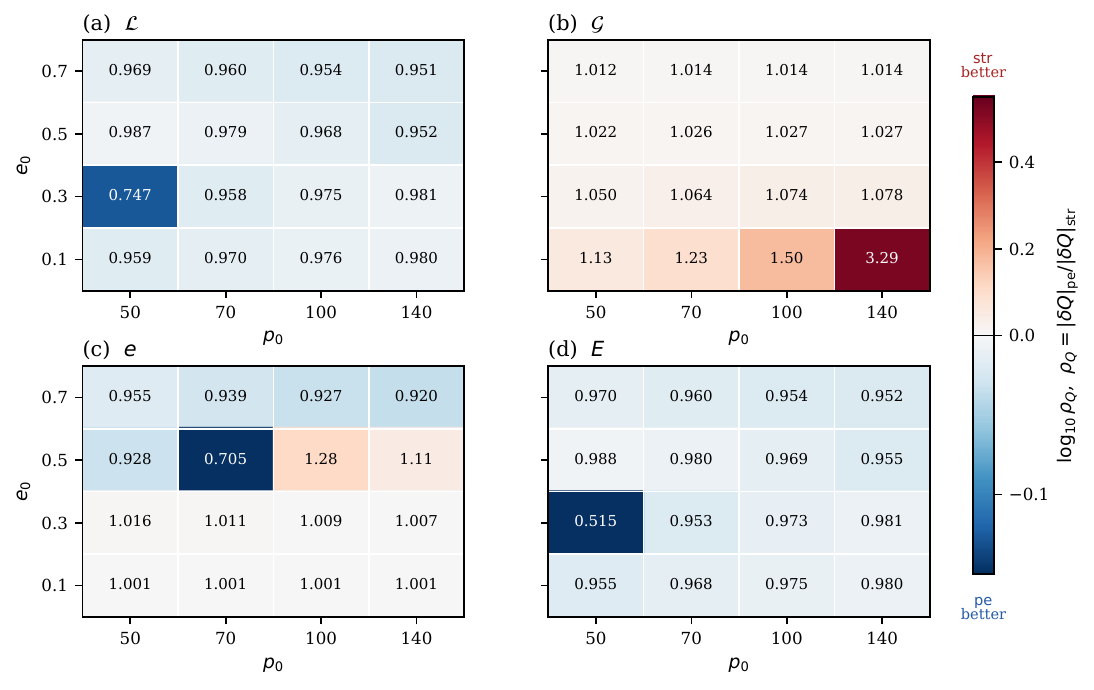}
	\caption{
		One-cycle comparison over the physical orbital-parameter grid
		\eqref{eq:orbital-parameter-grid}. Each cell reports the ratio
		\(\rho_Q=|\delta Q_1|_{\rm pe}/|\delta Q_1|_{\rm str}\) from
		Eq.~\eqref{eq:one-cycle-error-ratio}; blue cells
		(\(\rho_Q<1\)) favor the partially exponentiated map and red cells
		(\(\rho_Q>1\)) favor the strict truncation. Partial exponentiation
		improves \(\mathcal L\) and \(E\) throughout this grid, the strict map
		improves \(\mathcal G\), and the eccentricity ranking is mixed. The
		ratios should be read together with the normalized-error scale quoted
		in the text, since a large ratio can arise when both residuals are
		already small.
	}
	\label{fig:one-cycle-map-ratio-grid}
\end{figure*}

Our results are collected in Fig.~\ref{fig:one-cycle-map-ratio-grid}. For
\(\mathcal L\), the \({\rm pe}\) residual is smaller at all
16 points, with \(0.747\leq\rho_{\mathcal L}\leq0.987\). The same is true for
the energy, with \(0.515\leq\rho_E\leq0.988\). In contrast, \({\rm str}\)
gives the smaller \(\mathcal G\) residual throughout the grid, with
\(1.012\leq\rho_{\mathcal G}\leq3.295\). The largest of these ratios occurs at
\((p_0,e_0)=(140,0.1)\), where both angular-action errors are already very
small:
\(\mathcal E_{\mathcal G}^{\rm str}=3.94\times10^{-5}\) and
\(\mathcal E_{\mathcal G}^{\rm pe}=1.30\times10^{-4}\).
The eccentricity comparison is mixed:
\({\rm pe}\) is better at six of the 16 points and
\(0.705\leq\rho_e\leq1.283\). Near \(e_0=0.1\), the two eccentricity
residuals agree to about one part in \(10^3\).

The absolute normalized errors provide the necessary scale for these ratios.
For both prescriptions,
\(\mathcal E_Q<0.118\) for
\(Q\in\{\mathcal L,\mathcal G,E\}\) over the whole grid. The larger maximum
\(\mathcal E_e\simeq0.55\) occurs for \(e_0=0.1\), where the direct
one-cycle eccentricity change is only \(4.57\times10^{-4}\); it is therefore
mainly an amplification by the normalization rather than a large absolute
offset.

The grid therefore does not support a global preference for either
prescription. Partial exponentiation improves the radial-action and energy
updates in this domain but increases the angular-action residual, while the
eccentricity reflects both action errors. Because successive iterations
evaluate the kick on the actions produced by the preceding step, this
one-cycle ordering need not persist over many cycles.
Equation~\eqref{eq:diagnostic-residual-linearization} likewise explains why
the derived \(e\) and \(E\) diagnostics can rank the maps differently.

\section{Conclusions}
\label{sec:conclusions}

We have constructed the one-cycle first Magnusian, linear in the
radiation-reaction interaction, for planar eccentric binaries through relative
1PN order. The calculation combines the 1PN conservative normal form and its
Jacobi transport with the 2.5PN and 3.5PN radiation-reaction forces. The
result is a closed-form covector in normal-form Delaunay variables, with an
explicit embedding into ADM phase space, valid for \(0<e<1\) without an
eccentricity expansion. Unlike orbit-averaged balance equations, this
four-component generator supplies a one-cycle phase-space update: its
angle-indexed components determine the action losses, while its
action-indexed components encode the phase information from which the
return-time and apsidal corrections are obtained. It is therefore useful for
stroboscopic evolutions in which both secular losses and orbital dephasing
must be tracked.

We have also constructed the projection from the fixed conservative endpoint
of the one-cycle Magnusian to the perturbed periastron return section. This
projection produces the first-order correction to the return time,
and induces an additional apsidal contribution at
$\mathcal O(\lambda\varepsilon)$. The latter is therefore required for a
consistent periastron return map through relative 1PN order.

A central structural result is the independence of the first one-cycle
Magnusian from all eight Iyer--Will gauge parameters through relative 1PN
radiation-reaction order. After converting the generalized force to the doubled-space force covector and placing every representative on the same ADM phase
space, we found that each gauge response is a periodic coboundary: its
endpoint contribution vanishes over one radial cycle. This mechanism removes
the gauge dependence not only from the energy and angular-momentum losses, but
also from the two components that encode the interaction-picture angle
sector entering the return-time and apsidal projections.

Several independent checks support the closed-form result. The action changes
agree with direct mechanical work and torque, recover the standard circular
energy and angular-momentum fluxes, and reproduce the secular evolution of the
1PN radial quasi-Keplerian elements for \(0<e<1\), without an eccentricity
expansion. The angle-sector components satisfy the midpoint-shear identity
implied by the normal-form Jacobi propagator. Finally, the return-section
timing and apsidal corrections have been tested directly against the
numerical ADM evolution.

We also compared the strict and partially exponentiated uses of the first
Magnusian. Neither prescription is uniformly superior. For the representative
40-cycle evolution, \({\rm str}\) gives smaller state-space residuals at fixed
periastron index and the smaller first-cycle timing residual, whereas
\({\rm pe}_2\) and \({\rm pe}_3\) reduce the accumulated timing residual by
factors of about \(12\), with \({\rm pe}_2\) being marginally better than
\({\rm pe}_3\). Across the equal-mass one-cycle grid, partial exponentiation
improves the \(\mathcal L\) and \(E\) updates, the strict map improves
\(\mathcal G\), and the eccentricity ranking is mixed. State-space and
physical-time comparisons therefore probe complementary aspects of the map.
Partial exponentiation should be regarded as an observable-dependent
resummation, rather than as a systematic increase in perturbative accuracy.


The natural next step is the calculation of the independent second
Magnusian, and hence of \(\mathcal D_2\), at a specified, consistent PN
accuracy. This would complete the map through \(\mathcal O(\lambda^2)\) at
that accuracy, separate genuine second-order time-ordering effects from the
nested action of \(\mathcal D_1\), and turn the numerical indications found
here into a controlled higher-order construction. Further extensions
include higher-PN conservative and radiation-reaction input, spin effects, and
hereditary or other nonlocal interactions
\cite{Galley:2015kus,Damour:2014jta,Blanco:2024fte}. Related near-identity averaging methods have proved effective for
long-duration self-forced inspirals, including eccentric ones \cite{VanDeMeent:2018cgn,Lynch:2021ogr}. More broadly, one-cycle phase-space maps in canonical coordinates may provide
an efficient stroboscopic description of long eccentric inspirals while
retaining both the secular action losses and the dissipative phase
information.

\acknowledgments
We thank Francisco M. Blanco for helpful exchanges and for suggesting the physical-time comparison, and Elisa Grilli, Marta Orselli, and Gianluca Grignani for carefully reading the manuscript and providing valuable comments.
A.P. acknowledges financial support from
the Italian Ministry of University and Research (MUR)
through the program ``Dipartimenti di Eccellenza 2018--2022''
(Grant SUPER-C).

\clearpage
\appendix

\section{Periodic-coboundary form of the gauge response}
\label{app:periodic-coboundary}
The construction of Sec.~\ref{subsec:gauge-response} generates each of the
32 gauge responses directly from the complete Iyer--Will force and the
canonical conversion of Sec.~\ref{subsec:canonical-iyer-will-force}.
We exhibit here the property responsible for their vanishing integrals:
every rational response \(\widehat{\mathcal R}_{IA}(x)\) is an exact
derivative whose primitive vanishes at both endpoints. Throughout this appendix we suppress the common overall factor \(\lambda\)
in the response functions, since it does not affect any of its content.

After putting each nonzero response over a common denominator, it takes the
form
\begin{align}
	\widehat{\mathcal R}_{IA}(x)
	&=
	\frac{1}{\left(\sigma_-+\sigma_+ x^2\right)^n}
	\sum_{k=0}^{n-1}c_k^{IA}x^{2k},
	\\
	\sigma_-&=1-e,
	\qquad
	\sigma_+=1+e,
	\label{eq:gauge-rational-family}
\end{align}
where the denominator order \(n\) may depend on \(I\) and \(A\). In the row
ordering $(\alpha,\beta,\delta_1,\ldots,\delta_5,\epsilon_5)$ and the column ordering
$(\ell,g,\mathcal L,\mathcal G)$, the orders found by exact reduction are
\begin{equation}
	\mathsf N
	=
	\begin{pmatrix}
		7&6&7&5\\
		8&0&8&6\\
		6&5&6&0\\
		7&0&7&0\\
		6&5&6&0\\
		8&0&8&0\\
		7&0&7&0\\
		0&6&0&0
	\end{pmatrix}.
	\label{eq:gauge-denominator-order-matrix}
\end{equation}
Here a zero denotes an identically vanishing paired response, for which we set
$\mathcal B_{IA}=0$. Thus, 12 responses vanish already after pairing, while
the remaining 20 have $n\in\{5,6,7,8\}$. For each nonzero response, we
construct a candidate primitive of the form
\begin{equation}
	\mathcal B_{IA}(x)
	=
	\frac{x}{
		\left(\sigma_-+\sigma_+ x^2\right)^{n-1}
	}
	\sum_{k=0}^{n-2}
	b_k^{IA}x^{2k}.
	\label{eq:gauge-coboundary-primitive}
\end{equation}
Requiring
\begin{equation}
	\frac{\diff\mathcal B_{IA}}{\diff x}
	=
	\widehat{\mathcal R}_{IA}(x)
	\label{eq:coboundary-derivative-condition}
\end{equation}
gives the coefficient relation
\begin{equation}
	c_k^{IA}
	=
	\sigma_-(2k+1)b_k^{IA}
	+
	\sigma_+(2k-2n+1)b_{k-1}^{IA},
	\label{eq:coboundary-recurrence}
\end{equation}
with
\begin{equation}
	b_{-1}^{IA}=b_{n-1}^{IA}=0.
\end{equation}
The coefficients of the primitive are therefore generated recursively by
\begin{align}
	&b_0^{IA}
	=
	\frac{c_0^{IA}}{\sigma_-},
	\\
	&b_k^{IA}
	=
	\frac{c_k^{IA}}
	{\sigma_-(2k+1)}
	\nonumber\\
	&\quad
	-
	\frac{\sigma_+(2k-2n+1)b_{k-1}^{IA}}
	{\sigma_-(2k+1)},
	\qquad
	1\leq k\leq n-2.
	\label{eq:coboundary-coefficients}
\end{align}
The last numerator coefficient is not used in the recurrence. Instead, it
must satisfy the closure condition
\begin{equation}
	c_{n-1}^{IA}
	+
	\sigma_+\,b_{n-2}^{IA}
	=
	0.
	\label{eq:coboundary-closure}
\end{equation}

To illustrate how the preceding construction applies to the actual
gauge-response functions, consider the response of the \(\ell\) component to
the parameter \(\delta_4\). Its denominator order is \(n=8\), and exact
reduction gives
\begin{equation}
	\widehat{\mathcal R}_{\delta_4\ell}(x)
	=
	\frac{512e^5\nu}{5\mathcal L^6}
	\frac{
		x^4(1+x^2)
		\left[
		5\sigma_-
		-18ex^2
		-5\sigma_+x^4
		\right]
	}{
		\left(
		\sigma_-+\sigma_+x^2
		\right)^8
	}.
	\label{eq:representative-gauge-response}
\end{equation}
In the notation of Eq.~\eqref{eq:gauge-rational-family}, its numerator
coefficients are
\begin{subequations}
	\label{eq:representative-gauge-numerator}
	\begin{align}
		c_0^{\delta_4\ell}
		&=
		c_1^{\delta_4\ell}
		=
		c_6^{\delta_4\ell}
		=
		c_7^{\delta_4\ell}
		=
		0,
		\\
		c_2^{\delta_4\ell}
		&=
		\frac{512e^5\nu}{\mathcal L^6}\,
		\sigma_-,
		\\
		c_3^{\delta_4\ell}
		&=
		\frac{512e^5\nu}{5\mathcal L^6}
		\left(5-23e\right),
		\\
		c_4^{\delta_4\ell}
		&=
		-\frac{512e^5\nu}{5\mathcal L^6}
		\left(5+23e\right),
		\\
		c_5^{\delta_4\ell}
		&=
		-\frac{512e^5\nu}{\mathcal L^6}\,
		\sigma_+.
	\end{align}
\end{subequations}
The recurrence in Eq.~\eqref{eq:coboundary-coefficients} gives
\begin{subequations}
	\label{eq:representative-gauge-primitive-coefficients}
	\begin{align}
		b_0^{\delta_4\ell}
		&=
		b_1^{\delta_4\ell}
		=
		b_5^{\delta_4\ell}
		=
		b_6^{\delta_4\ell}
		=
		0,
		\\
		b_2^{\delta_4\ell}
		&=
		b_4^{\delta_4\ell}
		=
		\frac{512e^5\nu}{5\mathcal L^6},
		\\
		b_3^{\delta_4\ell}
		&=
		\frac{1024e^5\nu}{5\mathcal L^6},
	\end{align}
\end{subequations}
and hence
\begin{equation}
	\mathcal B_{\delta_4\ell}(x)
	=
	\frac{512e^5\nu}{5\mathcal L^6}
	\frac{
		x^5(1+x^2)^2
	}{
		\left(
		\sigma_-+\sigma_+x^2
		\right)^7
	}.
	\label{eq:representative-gauge-primitive}
\end{equation}
Direct differentiation and endpoint evaluation give
\begin{equation}
	\frac{\diff\mathcal B_{\delta_4\ell}}{\diff x}
	=
	\widehat{\mathcal R}_{\delta_4\ell}(x),
	\qquad
	\mathcal B_{\delta_4\ell}(0)
	=
	\lim_{x\to\infty}
	\mathcal B_{\delta_4\ell}(x)
	=
	0.
	\label{eq:representative-gauge-check}
\end{equation}
Therefore,
\begin{equation}
	\frac{\partial
		\chi_{{\rm 1PN},\ell}^{(1)}
	}{
		\partial\delta_4
	}
	=
	\int_0^\infty\!\diff x\,
	\widehat{\mathcal R}_{\delta_4\ell}(x)
	=
	0,
\end{equation}
although the corresponding response integrand does not vanish pointwise.

The same procedure is applied to the remaining 19 nonzero gauge-component combinations. In each case, the coefficients \(c_k^{IA}\)
are extracted directly from the corresponding rational response
\(\widehat{\mathcal R}_{IA}\), and
\(b_0^{IA},\ldots,b_{n-2}^{IA}\) are then generated recursively from
\(c_0^{IA},\ldots,c_{n-2}^{IA}\). Since \(c_{n-1}^{IA}\) does not enter this
construction, Eq.~\eqref{eq:coboundary-closure} constitutes an independent
check of the existence of the endpoint-trivial primitive. Including the
representative example above, all 20 nonzero responses satisfy the following
three identities exactly:
\begin{align}
	c_{n-1}^{IA}
	+
	\sigma_+\,b_{n-2}^{IA}
	&=
	0,
	\label{eq:closure-check}
	\\
	\frac{\diff\mathcal B_{IA}}{\diff x}
	-
	\widehat{\mathcal R}_{IA}(x)
	&=
	0,
	\label{eq:primitive-derivative-check}
	\\
	\mathcal B_{IA}(0)
	=
	\mathcal B_{IA}(\infty)
	&=
	0.
	\label{eq:primitive-endpoint-check}
\end{align}
Together with the 12 responses that vanish identically after pairing, these
checks verify all 32 derivative identities and all 64 endpoint identities.
It follows component by component that
\begin{align}
	\frac{\partial\chi_{{\rm 1PN},A}^{(1)}}
	{\partial\gamma_I}
	&=
	\int_0^\infty\!\diff x\,
	\widehat{\mathcal R}_{IA}(x)
	\nonumber\\
	&=
	\int_0^\infty\!\diff x\,
	\frac{\diff\mathcal B_{IA}}{\diff x}
	\nonumber\\
	&=
	\mathcal B_{IA}(\infty)
	-
	\mathcal B_{IA}(0)
	=
	0.
	\label{eq:coboundary-gauge-proof}
\end{align}

The gauge-dependent pullback integrands are therefore not equal pointwise.
Rather, changing the Iyer--Will parameters adds an exact, endpoint-trivial
term to each component of the paired one-cycle integrand. This
periodic-coboundary property is the mechanism behind the independence of the
first one-cycle Magnusian from all eight Iyer--Will gauge parameters through
relative 1PN radiation-reaction order.

\section{Formal radiation-reaction-strength diagnostic}
\label{app:formal-lambda-diagnostic}

To identify the leading powers of the radiation-reaction strength in the
difference between each finite map and the direct evolution, we perform an
auxiliary scaling test at fixed initial orbital parameters. Here
\({\rm pe}\) denotes the three-bracket partially exponentiated map of
Eq.~\eqref{eq:blanco-partial-exponential}. In this test, \(\lambda\) is varied
independently of \(\varepsilon\) solely as a formal bookkeeping parameter.

We collect the four state-space residuals into
\begin{equation}
	\bm R_n
	\equiv
	\left(
	\delta\mathcal L_n,
	\delta\mathcal G_n,
	\delta e_n,
	\delta E_n
	\right).
	\label{eq:numerical-residual-vector}
\end{equation}
For either prescription, the residual at fixed \(\lambda\) is expanded as
\begin{equation}
	\bm R_n(\lambda,\varepsilon)
	=
	\bm R_n^{[0]}(\lambda)
	+
	\varepsilon\bm R_n^{[1]}(\lambda)
	+
	\mathcal O(\varepsilon^2).
	\label{eq:residual-pn-expansion}
\end{equation}
The coefficient \(\bm R_n^{[1]}\) is extracted by centered differences at
\(h_1=1/100\), \(h_2=1/200\), and \(h_3=1/400\), rematching the initial ADM
state in every auxiliary evolution. Two-level Richardson extrapolation
\cite{Richardson_1927} gives
\begin{align}
	\bm R_{n,h}^{[1]}(\lambda)
	&=
	\frac{
		\bm R_n(\lambda,h)
		-
		\bm R_n(\lambda,-h)
	}{2h},
	\nonumber\\
	\bm R_{n,12}^{[1]}(\lambda)
	&=
	\frac{
		4\bm R_{n,h_2}^{[1]}(\lambda)
		-
		\bm R_{n,h_1}^{[1]}(\lambda)
	}{3},
	\nonumber\\
	\bm R_{n,23}^{[1]}(\lambda)
	&=
	\frac{
		4\bm R_{n,h_3}^{[1]}(\lambda)
		-
		\bm R_{n,h_2}^{[1]}(\lambda)
	}{3}.
	\label{eq:richardson-pn-projection}
\end{align}
The componentwise error estimate for the finer extrapolant is
\begin{equation}
	\bm\eta_n(\lambda)
	=
	\frac{1}{15}
	\left|
	\bm R_{n,23}^{[1]}(\lambda)
	-
	\bm R_{n,12}^{[1]}(\lambda)
	\right|.
	\label{eq:richardson-pn-error-estimate}
\end{equation}
The residual reconstructed through relative 1PN order and evaluated at the
physical bookkeeping value \(\varepsilon=1\) is therefore
\begin{equation}
	\bm R_{n,\leq{\rm 1PN}}(\lambda;\varepsilon=1)
	=
	\bm R_n(\lambda,0)
	+
	\bm R_{n,23}^{[1]}(\lambda).
	\label{eq:projected-nlo-residual}
\end{equation}

The same scalar projection is applied to the timing residual
\(\delta t_n\) in Sec.~\ref{subsec:physical-time-dephasing}. There it is
evaluated at \(\lambda=1\) over 40 cycles and does not enter the formal
strength-scaling analysis below.

For the orbit in Eq.~\eqref{eq:numerical-checkpoint}, evolved for 20 cycles
at \(\lambda\in\{1/2,3/4,1\}\), we define
\begin{equation}
	\Delta_Q^X(\lambda)
	\equiv
	\max_{0\leq n\leq20}
	\left|
	\delta Q_{n,\leq{\rm 1PN}}^X(\lambda)
	\right|,
	\qquad
	X\in\{{\rm str},{\rm pe}\}.
	\label{eq:formal-strength-residual}
\end{equation}
We fit
\(\Delta_Q^X\propto\lambda^{q_Q^X}\) by an unweighted linear regression of
\(\log\Delta_Q^X\) against \(\log\lambda\). In the component ordering
\((\mathcal L,\mathcal G,e,E)\), the fitted powers are
\begin{align}
	\bm q_{\rm str}
	&=
	(2.122,\,2.066,\,1.983,\,2.157),
	\nonumber\\
	\bm q_{\rm pe}
	&=
	(3.048,\,3.088,\,3.023,\,3.092).
	\label{eq:formal-scaling-orders}
\end{align}
The adjacent-interval slopes range from \(1.981\) to \(2.190\) for
\({\rm str}\) and from \(2.817\) to \(3.349\) for \({\rm pe}\). The \(\lambda=1/4\) data are excluded from the fit because, for several
components, the projected residual is no longer sufficiently well resolved
relative to the Richardson error estimate in
Eq.~\eqref{eq:richardson-pn-error-estimate} to provide a reliable scaling
point.

At \(\lambda=1\), the ratios of the projected residuals are
\begin{widetext}
\begin{equation}
	\left.
	\frac{\Delta_Q^{\rm pe}}
	{\Delta_Q^{\rm str}}
	\right|_{\lambda=1}
	=
	(2.76\times10^{-3},\,
	7.89\times10^{-4},\,
	3.88\times10^{-3},\,
	2.63\times10^{-3}),
\end{equation}
\end{widetext}
again in the ordering
\((\mathcal L,\mathcal G,e,E)\). Across the three sampled strengths, the
corresponding reduction factors range from approximately
\(2.6\times10^2\) to \(2.6\times10^3\). Every projected \({\rm pe}\) residual
remains more than 21 times larger than its estimated differencing error.

Over the sampled formal strengths, the residuals are consistent with
approximately quadratic scaling for \({\rm str}\) and cubic scaling for
\({\rm pe}\). For this orbit and projected diagnostic, the retained
\(\mathcal D_1^2/2\) term therefore removes nearly all of the resolved
quadratic difference between the strict map and the direct evolution. This
suggests that the independent \(\mathcal D_2\) contribution is subdominant
over the sampled range, but no general conclusion can be drawn for other
orbits or diagnostics.
\bibliography{bibliography}

\end{document}